\documentclass[prd, onecolumn]{revtex4}

\usepackage{graphicx}
\usepackage{dcolumn}
\usepackage{bm}
\usepackage{wrapfig}
\usepackage{epsfig}
\usepackage{breqn}
\usepackage{amssymb}
\usepackage{subfigure}

\begin{document}


\title{The Future Asymptotic Behaviour of a Non-Tilted Bianchi Type IV Viscous Model}

\author{Ikjyot Singh Kohli}
	\email{isk@yorku.ca}
\affiliation{York University - Department of Physics and Astronomy}
\author{Michael C. Haslam}
\email{mchaslam@mathstat.yorku.ca}
\affiliation{
York University - Department of Mathematics and Statistics
}

\date{\today}

\begin{abstract}
The future asymptotic behaviour of a non-titled Bianchi Type IV viscous fluid model is analyzed. In particular, we consider the case of a viscous fluid without heat conduction, and constant expansion-normalized bulk and shear viscosity coefficients. We show using dynamical systems theory that the only future attracting equilibrium points are the flat Friedmann-LeMaitre (FL) solution, the open FL solution and the isotropic Milne universe solution. We also show the bifurcations exist with respect to an increasing expansion-normalized bulk viscosity coefficient. It is finally shown through an extensive numerical analysis, that the dynamical system isotropizes at late times.
\end{abstract}
\maketitle 

\section{Introduction}
Spatially homogeneous and anisotropic models of the universe have undergone great study and continue to remain amongst the most popular areas of research in cosmology. Early-universe cosmological models for the most part have assumed the universe to be spatially homogeneous and anisotropic, with the important exception being the case of inhomogenous models such as the LeMaitre-Tolman-Bondi and ``Swiss-cheese'' models. However, if one begins with the idea of the early universe being spatially homogeneous and anisotropic, then to transition to the present-day Friedmann-LeMaitre-Robertson-Walker models requires the anisotropy in the former models to decay. The process by which this anisotropic decay occurs is arguably the most fundamental property of any early-universe model which aims to transition to the present-day models. For example, Belinskii, Khalatnikov, and Lifshitz \cite{lifshitz} studied the oscillatory approach to a singular point in relativistic cosmologies. Misner \cite{misnerc2} studied the anisotropic decay of the vacuum Bianchi Type IX/Mixmaster models of the universe. A very general approach to describing the isotropization of Bianchi models was described by Salucci and Fabbri \cite{salucci}. Coley and van den Hoogen \cite{vdh} studied causal anisotropic viscous fluid models and described conditions for such models to isotropize.  As for very recent work on this subject, Pradhan, Rai, and Singh \cite{pradhan} studied the Bianchi Type V bulk viscous models and showed that such models do isotropize for specific functional forms of the anisotropic scale factors.

Viscous models have become of general interest in early-universe cosmologies largely in two contexts. Gr{\o}n and Hervik (Chapter 13, \cite{hervik}) discuss these in some detail. The first relates through the idea of inflation through bulk viscosity. In models where bulk viscous terms are permitted to dominate, they drive the universe into a de Sitter-like state. Because of these processes, the models isotropize indirectly through the massive expansion. Shear viscosity is found to play an important role in universe models with dissipative fluids. The dissipative processes that result from shear viscous terms are thought to be highly effective during the early stages of the universe. In particular, neutrino viscosity is considered to be one of the most important factors in the isotropization of our universe.

As for Bianchi Type IV models specifically, Hervik, van den Hoogen, and Coley \cite{hervikvan} studied future asymptotic behaviour of tilted vacuum Bianchi Type IV models, and found that such models do not necessarily isotropize at late times. Uggla and Rosquist \cite{uggla} studied the orthogonal Bianchi Type IV model near the initial singularity with a vacuum or perfect-fluid source. We chose to study the isotropization behaviour of the Bianchi Type IV viscous model largely because such a study has not been taken on extensively in the literature, and perhaps, such a study will add to the already rich landscape of spatially homogeneous and anisotropic models of the early universe. 

We will use the Hubble-normalized dynamical systems approach based upon the theory of orthonormal frames pioneered by Ellis and MacCallum \cite{ellismac}, which reduces the Einstein field equations, a coupled set of ten hyperbolic nonlinear partial differential equations to a system of autonomous nonlinear first-order ordinary differential equations. We will also provide a fixed-point analysis of the dynamical system and make connections with the global dynamics through sophisticated numerical experiments.

\section{The Energy-Momentum Tensor for a Viscous Fluid}
In this section, we will derive the form of the energy-momentum tensor under concern, namely, for that of a viscous fluid without heat conduction. Recall that the energy-momentum tensor for a perfect fluid takes the form
\begin{equation}
T^{ab} = (\mu + p)u^{a}u^{b} - u^{c}u_{c} g^{ab} p.
\end{equation}
For the moment, letting $\mu + p = \mathcal{W}$, we obtain
\begin{equation}
\label{eq:enmom1}
T^{ab} = \mathcal{W} u^{a} u^{b} - u^{c}u_{c}g^{ab} p,
\end{equation}
Denoting the viscous contributions by $\mathcal{V}_{ab}$, we seek a modification of Eq. (\ref{eq:enmom1}) such that
\begin{equation}
\label{eq:enmom2}
T_{ab} = \mathcal{W}u_{a}u_{b} - u_{c}u^{c}g_{ab}p + \mathcal{V}_{ab}.
\end{equation}
To obtain the form of this additional tensor term, we note that from classical fluid mechanics, the Euler equation is given as
\begin{equation}
\left(\rho u_i \right)_{,t} = -\Pi_{ik,k},
\end{equation}
where $\Pi_{ik}$ is the momentum flux tensor. Also, recall that for a non-viscous fluid, one has the fundamental relationship
\begin{equation}
\label{eq:Pi1}
\Pi_{ik} = p \delta_{ik} + \rho u_i u_k.
\end{equation}
We simply add a term to Eq. (\ref{eq:Pi1}) that represents the viscous momentum flux, $\tilde{\Sigma}_{ik}$, to obtain 
\begin{equation}
\Pi_{ik} = p \delta_{ik} + \rho u_i u_k - \tilde{\Sigma}_{ik} = -S_{ik} + \rho u_i u_k.
\end{equation}
It is important to note that
\begin{equation}
S_{ik} = -p \delta_{ik} + \tilde{\Sigma}_{ik}
\end{equation}
is the stress tensor, while, $\tilde{\Sigma}_{ik}$ is the \emph{viscous} stress tensor. Note that, in what follows below, the viscous stress tensor, $\tilde{\Sigma}_{ik}$, is not to be confused with $\Sigma_{ik}$, the Hubble-normalized shear tensor.
The general form of the viscous stress tensor can be formed by recalling that viscosity is formed when the fluid particles move with respect to each other at different velocities, so this stress tensor can only depend on spatial components of the fluid velocity. We assume that these gradients in the velocity are small, so that the momentum tensor only depends on the first derivatives of the velocity in some Taylor series expansion. Therefore, $ \tilde{\Sigma}_{ik}$ is some function of the $u_{i,k}$.  In addition, when the fluid is in rotation, no internal motions of particles can be occurring, so we consider linear combinations of $u_{i,k} + u_{k,i}$, which clearly vanish for a fluid in rotation with some angular velocity, $\Omega_i$. The most general viscous tensor that can be formed is given by
\begin{equation}
\label{eq:sigma'}
\tilde{\Sigma}_{ik} = \eta \left(u_{i,k} + u_{k,i} - \frac{2}{3}\delta_{ik} u_{l,l}\right) + \xi \delta_{ik} u_{l,l},
\end{equation}
where $\eta$ and $\xi$ are the coefficients of shear and bulk/second viscosity, respectively \cite{landau_fluid} \cite{kundu}. In Eq. (\ref{eq:sigma'}), we note that $\delta_{ik} u_{l,l}$  is an expansion rate tensor, and $\left(u_{i,k} + u_{k,i} - \frac{2}{3}\delta_{ik} u_{l,l}\right)$ represents the shear rate tensor. Since we would like to generalize this expression to the general relativistic case, we replace the partial derivatives above with covariant derivatives, and the Kroenecker tensor with a more general metric tensor, that is, $\delta_{ik} \rightarrow g_{ik}$. We thus have that
\begin{equation}
\label{eq:sigma'2}
\tilde{\Sigma}_{ik} = \eta \left(u_{i;k} + u_{k;i} - \frac{2}{3}g_{ik} u_{l;l}\right) + \xi g_{ik} u_{l;l}.
\end{equation}
Denoting the shear rate tensor as $\sigma_{ab}$, and the expansion rate scalar as $\theta \equiv u^{a}_{;a}$, Eq. (\ref{eq:sigma'2}) becomes
\begin{equation}
\label{eq:vab}
\mathcal{V}_{ab} = -2\eta \sigma_{ab} - \xi \theta h_{ab}.
\end{equation}
Since we are interested in the Hubble-normalized approach, we will make use of the definition $\theta \equiv 3H$, where $H$ is the Hubble parameter. This means that Eq. (\ref{eq:sigma'2}) becomes
\begin{equation}
\label{eq:vab2}
\mathcal{V}_{ab} = -2\eta \sigma_{ab} - 3\xi H h_{ab}.
\end{equation}
Substituting Eq. (\ref{eq:vab2}) into Eq. (\ref{eq:enmom2}) we finally obtain the required form of the energy-momentum tensor as
\begin{equation}
\label{eq:enmomfinal}
T_{ab} = \left(\mu + p\right)u_{a}u_{b} - u_{c}u^{c}g_{ab}p - 2\eta \sigma_{ab} - 3\xi H h_{ab}.
\end{equation}
For simplicity, we shall let $\pi_{ab} = -2\eta \sigma_{ab}$ denote the anisotropic stress tensor, and commit to the metric signature $(-1,+1,+1,+1)$ such that Eq. (\ref{eq:enmomfinal}) takes the form
\begin{equation}
\label{eq:enmomfinal2}
T_{ab} = \left(\mu + p\right)u_{a}u_{b} + g_{ab}p  - 3\xi H h_{ab} + \pi_{ab}.
\end{equation}

\section{Bianchi Type IV Universe Dynamics}
With the required energy-momentum tensor in hand, we will now derive the Bianchi Type IV dynamical equations. The general evolution equations for any Bianchi type have already been derived in \cite{hewittbridsonwainwright}, \cite{herviklim} and \cite{elliscosmo}, and we will simply make use of their results in this section. 

The general evolution equations in the expansion-normalized variables are
\begin{eqnarray}
\label{eq:evolutionsys1}
\Sigma_{ij}' &=& -(2-q)\Sigma_{ij} + 2\epsilon^{km}_{(i}\Sigma_{j)k}R_{m} - \mathcal{S}_{ij} + \Pi_{ij} \nonumber \\
N_{ij}' &=& qN_{ij} + 2\Sigma_{(i}^{k}N_{j)k} + 2 \epsilon^{km}_{(i}N_{j)k}R_{m} \nonumber \\
A_{i}' &=& qA_{i} - \Sigma^{j}_{i}A_{j} + \epsilon_{i}^{km}A_{k} R_{m}\nonumber \\
\Omega' &=& (2q - 1)\Omega - 3P - \frac{1}{3}\Sigma^{j}_{i}\Pi^{i}_{j} + \frac{2}{3}A_{i}Q^{i} \nonumber \\
Q_{i}' &=& 2(q-1)Q_{i} - \Sigma_{i}^{j}Q_{j} - \epsilon_{i}^{km}R_{k}Q_{m} + 3A^{j}\Pi_{ij} + \epsilon_{i}^{km}N_{k}^{j}\Pi_{jm}.
\end{eqnarray}
These equations are subject to the constraints
\begin{eqnarray}
\label{eq:constraints1}
N_{i}^{j}A_{j} &=& 0 \nonumber \\
\Omega &=& 1 - \Sigma^2 - K \nonumber \\
Q_{i} &=& 3\Sigma_{i}^{k} A_{k} - \epsilon_{i}^{km}\Sigma^{j}_{k}N_{jm}.
\end{eqnarray}
In Eqs. (\ref{eq:evolutionsys1}) and (\ref{eq:constraints1}) we have made use of the following notation:
\begin{equation}
\label{eq:notation1}
\left(\Sigma_{ij}, R^{i}, N^{ij}, A_{i}\right) = \frac{1}{H}\left(\sigma_{ij}, \Omega^{i}, n^{ij}, a_{i}\right) , \quad \left(\Omega, P, Q_{i}, \Pi_{ij}\right) = \frac{1}{3H^2}\left(\mu, p, q_{i}, \pi_{ij}\right).
\end{equation}
In the expansion-normalized approach, $\Sigma_{ab}$ denotes the kinematic shear tensor, and describes the anisotropy in the Hubble flow, $A_{i}$ and $N^{ij}$ describe the spatial curvature, while $\Omega^{i}$ describes the relative orientation of the shear and spatial curvature eigenframes.
In addition,  $\mu$ and $p$ denote the \emph{total} energy density and total effective pressure, and are found by evaluating
\begin{equation}
\label{eq:totaldensitypressuredef}
\mu = u^{a} u^{b} T_{ab}, \quad p = \frac{1}{3} h^{ab} T_{ab},
\end{equation}
where, $h_{ab} = u_{a}u_{b} + g_{ab}$ denotes the projection tensor, and $u^{a}$, the fluid four-velocity \cite{herviklim}.
Since we are interested in a \emph{non-tilted} cosmology, the fluid is taken to be geodesic and irrotational, and thus has four-velocity $u^{a} = (1,0,0,0)$. We first note that the total energy density is indeed just $\mu$ as can be seen by applying the definition above. In addition, the total effective pressure is found from Eqs. (\ref{eq:totaldensitypressuredef}) and (\ref{eq:enmomfinal2}) to be
\begin{eqnarray}
p = \frac{1}{3}h^{ab}T_{ab} =  \tilde{p} - 3\xi H,
\end{eqnarray}
where $\tilde{p}$ denotes the fluid pressure in the barotropic equation of state, such that  $\tilde{p} = w \mu$. This implies that
\begin{equation}
\label{eq:P1}
P = w\Omega - 3 \xi_{0},
\end{equation}
where we have defined the equation of state
\begin{equation}
\frac{\xi}{H} \equiv 3\xi_{0},
\end{equation}
with $\xi_{0}$ being a nonnegative constant.
Similarly, we find that
\begin{equation}
\Pi_{ab} = - 2\eta_{0} \Sigma_{ab},
\end{equation}
where $\eta_{0}$ is a nonnegative constant as defined by the equation of state
\begin{equation}
\frac{\eta}{H} = 3 \eta_{0}.
\end{equation}
From these definitions of the expansion-normalized shear and bulk viscosity parameters, $\eta_{0}, \xi_{0}$ we would like to stress that throughout the proceeding analysis, we consider these parameters to be \emph{nonnegative constants}.

Since the fluid four-velocity is taken to be $u^{a} = (1,0,0,0)$, the quantity $q_{a} = Q_{a}3H^2$ vanishes by definition:
\begin{eqnarray}
q^{a} \equiv -h^{b}_{a} u^{c} T_{bc} =  -\left(u_{a}u^{0} + \delta^{a}_{0}\right)T_{00} = 0.
\end{eqnarray}

Our dynamical system evolves according to a dimensionless time variable, $\tau$ such that
\begin{equation}
\frac{dt}{d\tau} = \frac{1}{H},
\end{equation}
where $H$ is the Hubble parameter.  The deceleration parameter $q$ is very important in the expansion-normalized approach, and through the evolution equation for $H$
\begin{equation}
H' = -(1 + q) H,
\end{equation}
one can show that $q$ is defined as
\begin{eqnarray}
\label{eq:q1}
q &\equiv& 2\Sigma^2 + \frac{1}{2}\left(\Omega + 3P\right) \nonumber \\
&=& \frac{1}{3}\left(\Sigma_{ab}\Sigma^{ab}\right) + \Omega \left[\frac{1}{2} + \frac{3}{2}w\right] - \frac{9}{2}\xi_{0},
\end{eqnarray}
where we have made use of Eq. (\ref{eq:P1}) and the definition $\Sigma^2 \equiv \frac{1}{6}\Sigma^{ab}\Sigma_{ab}$.

Following the convention in \cite{hervik}, for the Bianchi Type IV models, we have
\begin{equation}
\label{eq:biv}
A^{i} = A \delta^{i}_{3} \neq 0, \quad N_{11} \neq 0, \quad N_{22} = N_{33} = 0.
\end{equation}
Computing the evolution equations requires one to first compute the Hubble-normalized spatial curvature variables, $\mathcal{S}_{ij}$  and $K$. According to (A.7) in \cite{hewittbridsonwainwright}, we have that
\begin{equation}
\mathcal{S}_{ab} = B_{ab} - \frac{1}{3}B^{u}_{u}\delta_{ab} - 2 \epsilon^{uv}_{(a}N_{b)u}A_{v}, \quad K = \frac{1}{12}B^{u}_{u} + A_{u}A^{u},
\end{equation}
where $B_{ab} \equiv 2N^{u}_{a}N_{ub} - N^{u}_{u}N_{ab}$.
Evaluating these expressions for the Bianchi IV model, we obtain
\begin{eqnarray}
\label{eq:Scalcs}
\mathcal{S}_{11} = \frac{2}{3}N_{11}, \quad \mathcal{S}_{12} = N_{11}A,  \quad \mathcal{S}_{22} = \mathcal{S}_{33} = -\frac{N_{11}^2}{3}, \quad K = A^2 + \frac{N_{11}^2}{12}.
\end{eqnarray}
The constraints in Eq. (\ref{eq:constraints1}) imply that
\begin{equation}
\Sigma_{31} = \Sigma_{32} = 0, \quad 3A\Sigma_{33} + N_{11} \Sigma_{21} = 0,
\end{equation}
that is, that $\Sigma_{21} \neq 0$.
In addition, the $\Sigma_{13}'$ and $\Sigma_{23}'$ equations from Eqs. (\ref{eq:evolutionsys1}) imply that $R_{1} = R_{2} = 0$. Looking at the $N_{12}'$ equation from the same set implies that $R_{3} = \Sigma_{12}$. We have therefore uniquely determined $R_{i}$ in terms of $\Sigma_{ij}$, and can see that the independent expansion-normalized variables are: $\Sigma_{22}, \Sigma_{33}, \Sigma_{12}, A$, and $N_{11}$. Taking advantage of the fact that the shear tensor, $\Sigma_{ab}$ is trace-free, we will define new variables as follows:
\begin{eqnarray}
\Sigma_{+} &=& \left(\Sigma_{22} + \Sigma_{33}\right), \nonumber \\
\Sigma_{-} &=& \frac{1}{\sqrt{3}} \left(\Sigma_{33} - \Sigma_{22}\right), \nonumber \\
N_{1} &=& N_{11}, \nonumber \\
\Sigma_{3} &=& \frac{1}{\sqrt{3}} \Sigma_{12}.
\end{eqnarray}
As mentioned in \cite{hewittbridsonwainwright} and \cite{elliscosmo}, the off-diagonal shear component $\Sigma_{3}$ determines the angular velocity of the spatial frame. The set of independent expansion-normalized variables is then
\begin{equation}
\label{eq:set1}
\left(\Sigma_{+}, \Sigma_{-}, N_{1},  A, \Sigma_{3}\right).
\end{equation}

The evolution equations for these variables are
\begin{eqnarray}
\label{eq:systemfinal1}
\Sigma_{+}' &=& \frac{2N_{1}^2}{3} - 4\Sigma_{3}^2 + \Sigma_{+}\left[-2+q-2\eta_{0}\right],  \nonumber \\
\Sigma_{-}' &=&  \frac{4\Sigma_{3}^2}{\sqrt{3}} + \Sigma_{-}\left[q - 2\left(1+\eta_{0}\right)\right],   \nonumber \\
\Sigma_{3}' &=& \Sigma_{3} \left[-(2-q) + \frac{3}{2}\Sigma_{+} - \frac{\sqrt{3}}{2}\Sigma_{-} - 2 \eta_{0}\right] - \frac{1}{\sqrt{3}}N_{1}, \nonumber \\
N_{1}' &=& N_{1} \left[q - 2\Sigma_{+}\right], \nonumber \\
A' &=& A \left[q - \frac{\sqrt{3}\Sigma_{-}}{2} - \frac{\Sigma_{+}}{2}\right],
\end{eqnarray}
where
\begin{equation}
\label{eq:qdef}
q = 2 \left(\Sigma_{3}^2+\Sigma_{-}^2+\Sigma_{+}^2\right)-\frac{1}{24} \left[12 A^2+N_{1}^2+12 \left(-1+\Sigma_{3}^2+\Sigma_{-}^2+\Sigma_{+}^2\right)\right] (1+3 w)-\frac{9 \xi_{0} }{2},
\end{equation}
with the constraint given by
\begin{equation}
\label{eq:constraintfinal1}
g(\mathbf{x}) = \frac{3}{2}A \left(\sqrt{3}\Sigma_{-} + \Sigma_{+}\right) + \sqrt{3} N_{1}  \Sigma_{3} = 0.
\end{equation}
The state space is the subset of $\mathbb{R}^{5}$ defined by the physical inequality $\Omega \geq 0$, which is equivalent to
\begin{equation}
\Sigma_{+}^2 +\Sigma_{-}^2 + \Sigma_{3}^2 + A^2 + \frac{1}{12}N_{1}^2 \leq 1,
\end{equation}
This restriction indeed implies that the state space is bounded.  

We additionally find that the evolution equations (\ref{eq:systemfinal1}) have a transformation invariance such that
\begin{equation}
\left[\Sigma_{+}, \Sigma_{-}, \Sigma_{3}, N_{1}, A\right] \rightarrow \left[\Sigma_{+}, \Sigma_{-}, \Sigma_{3}, \pm N_{1}, \pm A\right].
\end{equation}
One can therefore assume without loss of generality that $N_{1} \geq 0$ and $A \geq 0$.

\section{A Local Stability Analysis}
In this section, we consider the local stability of the equilibrium points of the system (\ref{eq:systemfinal1}) and (\ref{eq:constraintfinal1}). The critical points are points $\mathbf{x} = \mathbf{a}$ that simultaneously satisfy
\begin{equation}
\mathbf{f}(\mathbf{a}) = 0, \quad g(\mathbf{a}) = 0,
\end{equation}
where $\mathbf{f}$ denotes the right-hand-side of the system (\ref{eq:systemfinal1}). The local stability is determined by linearizing the system  (\ref{eq:systemfinal1})  at $\mathbf{x} = \mathbf{a}$, which leads to the relationship $\mathbf{x}' = D \mathbf{f(a)} \mathbf{x}$. The stability of the system is then determined by finding the eigenvalues and eigenvectors of the derivative matrix $D \mathbf{f(a)}$. Because of the constraint, we are to only consider physical eigenvalues, that is, eigenvalues whose corresponding eigenvectors are orthogonal to to $\nabla g (\mathbf{a})$. The gradient of the constraint Eq. (\ref{eq:constraintfinal1}) is found to be
\begin{equation}
\nabla g(\mathbf{x}) = \left[\frac{3A}{2},  \frac{3\sqrt{3}}{2}A,  \sqrt{3}N_{1},  \sqrt{3}\Sigma_{3}, \frac{3}{2} \left(\sqrt{3}\Sigma_{-} + \Sigma_{+}\right)\right].
\end{equation}


\subsection{Equilibrium Point 1}
The first equilibrium point is found to be:
\begin{equation}
\left[\Sigma_{+}, \Sigma_{-}, \Sigma_{3}, N_{1}, A\right] = \left[0,0,0,0,0\right].
\end{equation}
The cosmological parameters at this point take the form:
\begin{equation}
\label{eq:cosmoparamsnew1}
\Omega = 1, \quad q = \frac{1}{2} \left(1 + 3w - 9\xi_{0}\right),  \quad \Sigma^2 = 0,
\end{equation}
where
\begin{equation}
\xi_{0} \geq 0, \quad \eta_{0} \geq 0, \quad -1 \leq w \leq 1.
\end{equation}
The eigenvalues corresponding to this critical point are:
\begin{equation}
\lambda_{1} = \lambda_{2} = \frac{1}{2} \left[1 + 3w - 9\xi_{0}\right], \quad \lambda_{3} = \lambda_{4} = \lambda_{5} = \frac{1}{2} \left[-3 + 3w - 4 \eta_{0} - 9\xi_{0}\right].
\end{equation}
This equilibrium point is a local sink (all of the eigenvalues have negative real parts) if $\eta_{0} \geq 0$, and
\begin{equation}
\label{eq:restr1}
\left\{\left[0 \leq \xi_{0} \leq \frac{4}{9}\right] \wedge \left[-1 \leq w < \frac{1}{3}\left(-1 + 9 \xi_{0}\right)\right]\right\} \vee \left\{\left[\xi_{0} > \frac{4}{9}\right] \wedge \left[ -1 \leq w < 1\right] \right\}.
\end{equation}

Based on the cosmological parameters (\ref{eq:cosmoparamsnew1}), we see that this equilibrium point represents a non-vacuum, flat Friedmann-LeMa\^{i}tre (FL) universe \cite{hewittbridsonwainwright} \cite{ellis}.  An important point to note is that $q = -1$ when $0 \leq \xi_{0} \leq \frac{2}{3}$ and $w = 3\xi_{0} - 1$, thus, the equilibrium point in the domain defined by these values of $\eta_{0} ,\xi_{0}$ ,and $w$ does not correspond to a self-similar solution. In particular, if one chooses $\xi_{0} = 0$ such that $w = -1$, the corresponding model is locally the de Sitter solution \cite{elliscosmo}.
\subsection{Equilibrium Point 2}
The second equilibrium point is found to be:
\begin{equation}
\label{eq:eqpoint2}
\left[\Sigma_{+}, \Sigma_{-}, \Sigma_{3}, N_{1}, A\right] = \left[0,0,0,0,\frac{\sqrt{1+3 w-9 \xi_{0} }}{\sqrt{1+3 w}}\right].
\end{equation}
The cosmological parameters at this point take the form:
\begin{equation}
\label{eq:cosmoparamsnew2}
\Omega = \frac{9 \xi_{0}}{1 + 3w}, \quad q = 0, \quad \Sigma^2 = 0.
\end{equation}
This equilibrium point represents a Bianchi Type V model if
\begin{equation}
\label{eq:restr4}
\eta_{0} \geq 0, \quad \left\{0 \leq \xi_{0} < \frac{4}{9}\right\} \wedge \left\{\frac{1}{3} \left[-1 + 9 \xi_{0}\right] < w < 1\right\}.
\end{equation}
The eigenvalues corresponding to this critical point are:
\begin{equation}
\label{eq:eigV}
\lambda_{1} = 0, \quad \lambda_{2} = \lambda_{3} = -2 \left(1 + \eta_{0}\right), \quad \lambda_{4} = -1 - 3w + 9 \xi_{0}.
\end{equation}
We see that this equilibrium point is a non-isolated equilibrium point in general, because $\lambda_{1} = 0$. Whether it is a sink or a source depends on the signs of the other eigenvalues. Note that, $\lambda_{2} = \lambda_{3} = -2(1 + \eta_{0}) < 0,  \quad \forall \mbox{  } \eta_{0} \geq 0$, so the equilibrium classification depends on $\lambda_{4}$ alone. In particular, 
\begin{equation}
\label{eq:restrV}
\lambda_{4} = -1 - 3w + 9 \xi_{0} < 0 \Leftrightarrow \left\{0 \leq \xi_{0} < \frac{4}{9}\right\} \wedge \left\{\frac{1}{3} \left[-1 + 9 \xi_{0}\right] < w < 1\right\}.
\end{equation}
Therefore, we see that, if and only if $\lambda_{4} < 0$, this equilibrium point is a local sink (Section 4.3.4 \cite{ellis}). One can also show that given the restrictions in (\ref{eq:restr4}), $\lambda_{4} \neq 0$, and $\lambda_{4} \not > 0$. 

An interesting point is that if one chooses $\xi_{0} = 0$, then $ A = 1$, $\Omega = q = \Sigma^2 = 0$, and this equilibrium point represents the Milne universe.


\subsection{Other Possible Equilibrium Points}
We should note that in addition to the equilibrium points found above, there are additional ones that are purely mathematical. We are forced to ignore these points on physical grounds, because in order for $\Omega \geq 0$, we would have to have either $N < 0$, $A < 0$, $\xi_{0} < 0$, $\eta_{0} < 0$, $w < -1$ or $w > 1$. The first pair violate the Bianchi Type IV requirements, the second pair violate the requirement that any fluid must have nonnegative viscosity coefficients, and the last pair violate the well-known equation of state restrictions in cosmology. As a particular example, a fluid having an equation of state for which $w > 1$ implies that the matter under consideration has the speed of sound exceeding the speed of light, which would violate relativity theory.

As argued by Hervik et. al. \cite{hervikvan}, since we are only concerned with the future asymptotic behaviour of the Bianchi Type IV model, we will not be concerned with Type I Vacuum equilibrium points for which $N_{1} = A = \Omega = 0$, since for these models all of the equilibrium points are Kasner circles of which none are stable in the future, that is, they all represent local sources \cite{hervikvan} \cite{ellis}.

\subsection{Bifurcation Behaviour}
The physical equilibria found above are related to each other by a sequence of bifurcations, which can be understood as follows. The linearizations of the equations for $N_{1}$ and $A$ at the flat FL point are
\begin{eqnarray}
N_{1}' &=& \frac{1}{2} \left(1 + 3w - 9 \xi_{0}\right) N_{1}, \nonumber \\
A' &=& \frac{1}{2} \left(1 + 3w - 9 \xi_{0}\right) A,
\end{eqnarray}
which show that $N_{1}$ and $A$ destabilize the flat FL point if $\xi_{0} = \frac{1}{9} \left(1 + 3w\right)$, and that there is a bifurcation from the Bianchi Type V / Open FL point to the Bianchi Type I / Flat FL point, which can been seen from the ranges of $\xi_{0}$ in Table I.  Further, the linearization of the $A'$ equation at the Bianchi Type V point is found to be
\begin{equation}
A' = \left(-\frac{\sqrt{1+3 w-9 \xi_{0} }}{2 \sqrt{1+3 w}}\right)\Sigma_{+} - \left(-\frac{\sqrt{3+9 w-27 \xi_{0} }}{2 \sqrt{1+3 w}}\right)\Sigma_{-} + \left(-1 - 3 w + 9 \xi_{0}\right)A,
\end{equation}
which destabilizes the Bianchi Type V point if $\xi_{0} = \frac{1}{9} \left(1 + 3w\right)$ and $-\frac{1}{3} < w < 1$. It is clear then that with respect to this analysis, the line
\begin{equation}
\xi_{0} = \frac{1}{9} \left(1 + 3w\right)
\end{equation}
is very important as it governs the bifurcations of the system.
We give in FIG. 1 a useful summary diagram of the bifurcation regions in terms of the viscosity coefficients.
\begin{figure}[h]
\label{fig:viscdiagram}
\caption{This figure gives a schematic view of the viscosity coefficients as related to specific Bianchi type regions. The large white arrows indicate bifurcation transitions in terms of increasing expansion-normalized bulk-viscosity coefficient, where $\eta_{0}$ is assumed to be nonnegative in general. Note how the different regions are bounded by the lines $w = \frac{1}{3}\left[-1 + 9\xi_{0}\right]$ and $ \xi_{0} = \frac{4}{9}$.  Also indicated on the diagram by a thick black line is the line $\xi_{0} = 0$ in the BV region which indicates Milne universe solutions. For completeness, we have included the line $w = -1 + 3 \xi_{0}$, which for $0 \leq \xi_{0} \leq \frac{2}{3}$, represents non-self-similar solutions. In particular, the point $\xi_{0} = 0, w = -1$ represents the de Sitter solution.}
\includegraphics{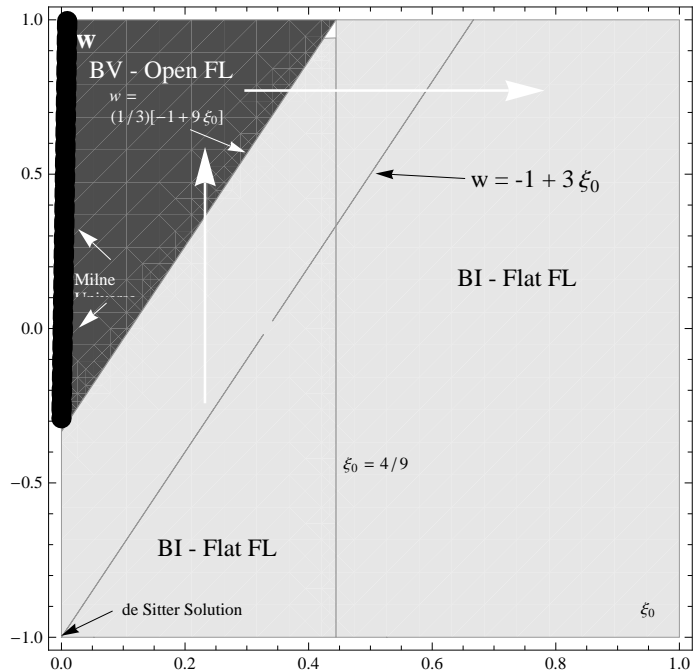}
\end{figure}
\newpage

\section{Late-Time Asymptotic Behaviour}
The goal of this section is to complement the preceding stability analysis of the equilibrium points with extensive numerical experiments in order to confirm that the local results are in fact global in nature.

By the Hartman-Grobman theorem for hyperbolic equilibrium points, and the invariant manifold theorem for non-isolated equilibrium points \cite{ellis},  we know that for any local sink, any orbit that enters a sufficiently small neighbourhood of the sink approaches the sink as $\tau \to \infty$ \cite{hewittbridsonwainwright}. The numerical solutions to the dynamical system presented below also provide strong evidence that for the given values of $\xi_{0}$ and $w$, the local sinks are the future attractors of the evolution equations. For each numerical solution, we chose the initial conditions such that the constraints (\ref{eq:constraints1}), (\ref{eq:constraintfinal1}) were satisfied, and are indicated by asterisks in the figures below. For completeness, we have listed the initial conditions used in Table I in the appendix.

Although numerical integrations were done from $0 \leq \tau \leq 1000$, for demonstration purposes, we presented solutions for shorter time intervals. We completed numerical integrations of the dynamical system for physically interesting cases of $w$ equal to 0 (dust), 0.325 (a dust/radiation mixture), and $\frac{1}{3}$ (radiation).

\newpage
\subsection{$\xi_{0} = \frac{4}{9}$, $\eta_{0} = 1$, $w = 0$ (Dust)}
\begin{figure}[h]
\label{fig:fig1}
\caption{This figure shows the dynamical system behaviour for $\xi_{0} = \frac{4}{9}$, $\eta_{0} = 1 $, and $w = 0$. The plus sign indicates the equilibrium point. Notice how the equilibrium point in this case, the Bianchi Type I / Flat FL point is indeed the local sink. The model also isotropizes as can be seen from the last figure, where $\Sigma_{\pm} \to 0$ as $\tau \to \infty$.}
\includegraphics*[scale = 0.45]{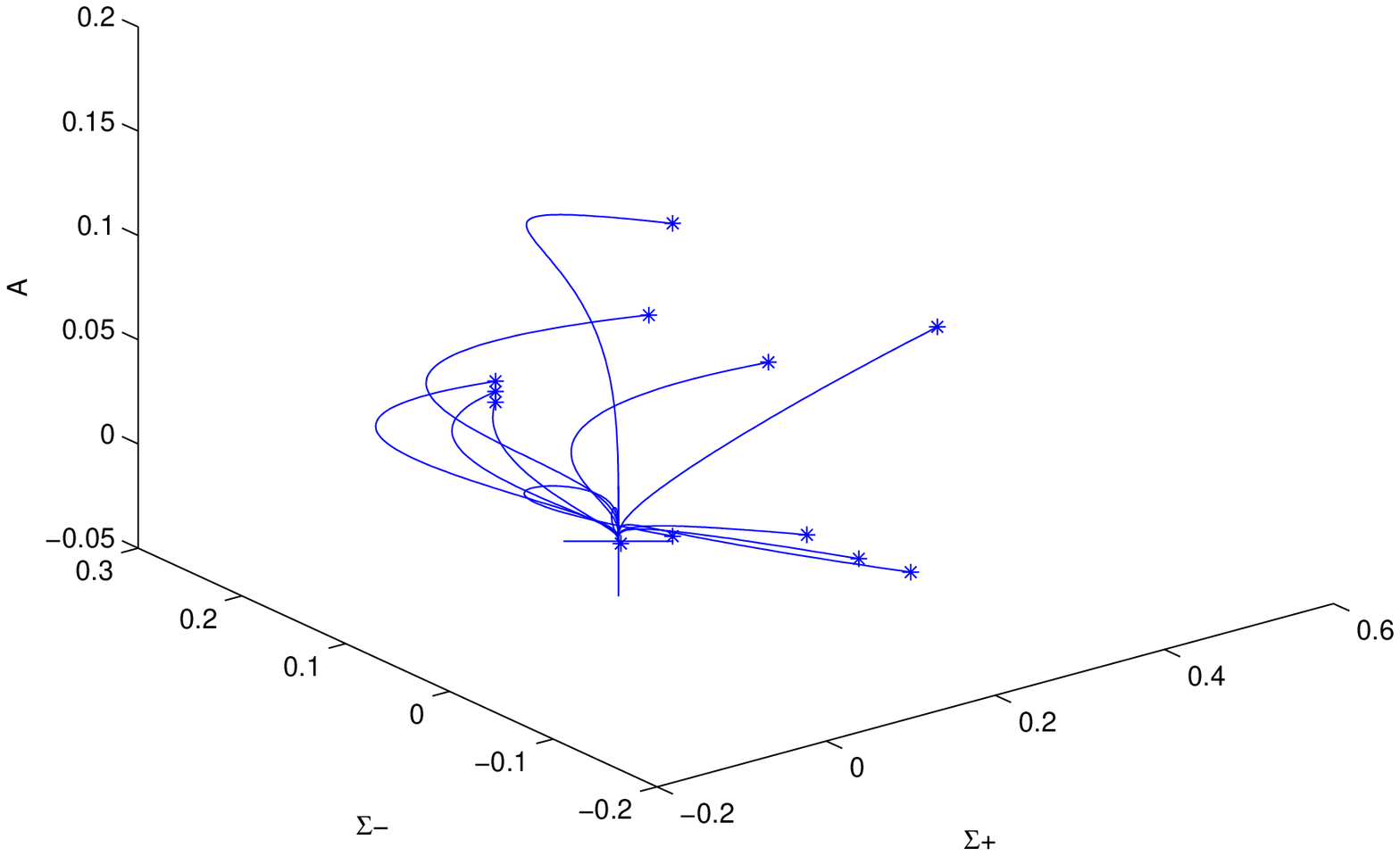}
\includegraphics*[scale = 0.45]{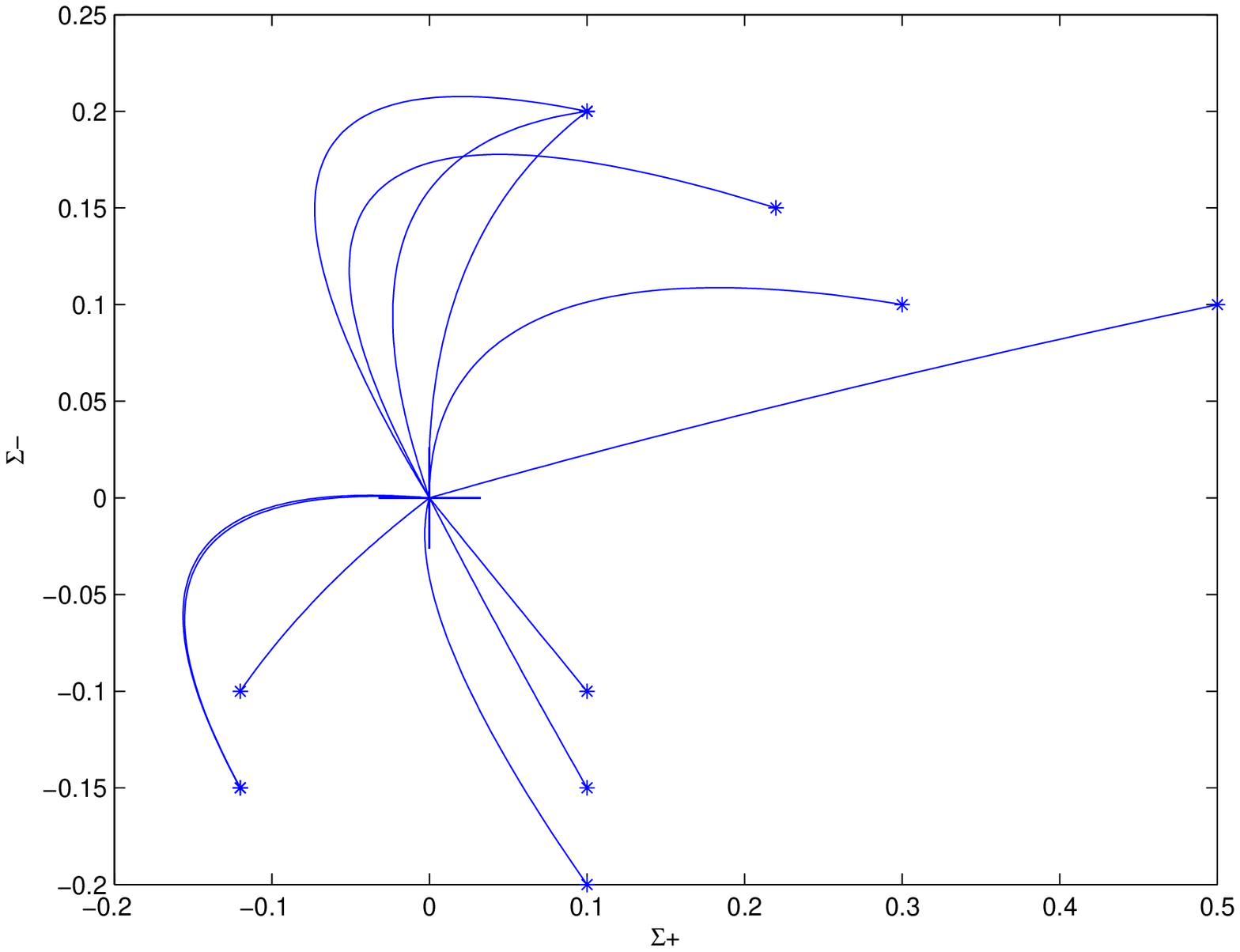}
\linebreak
\linebreak
\includegraphics*[scale = 0.45]{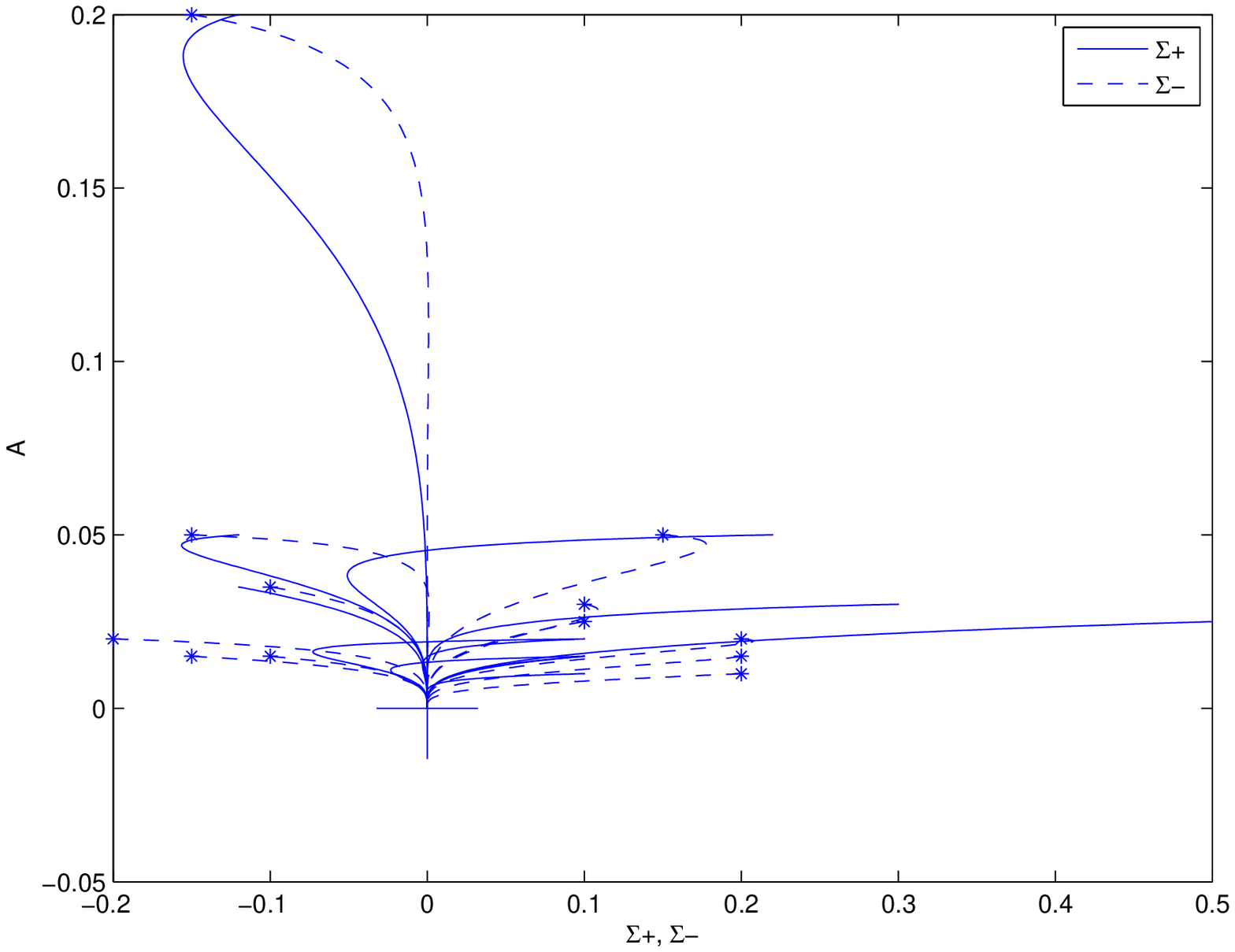}
\includegraphics*[scale = 0.45]{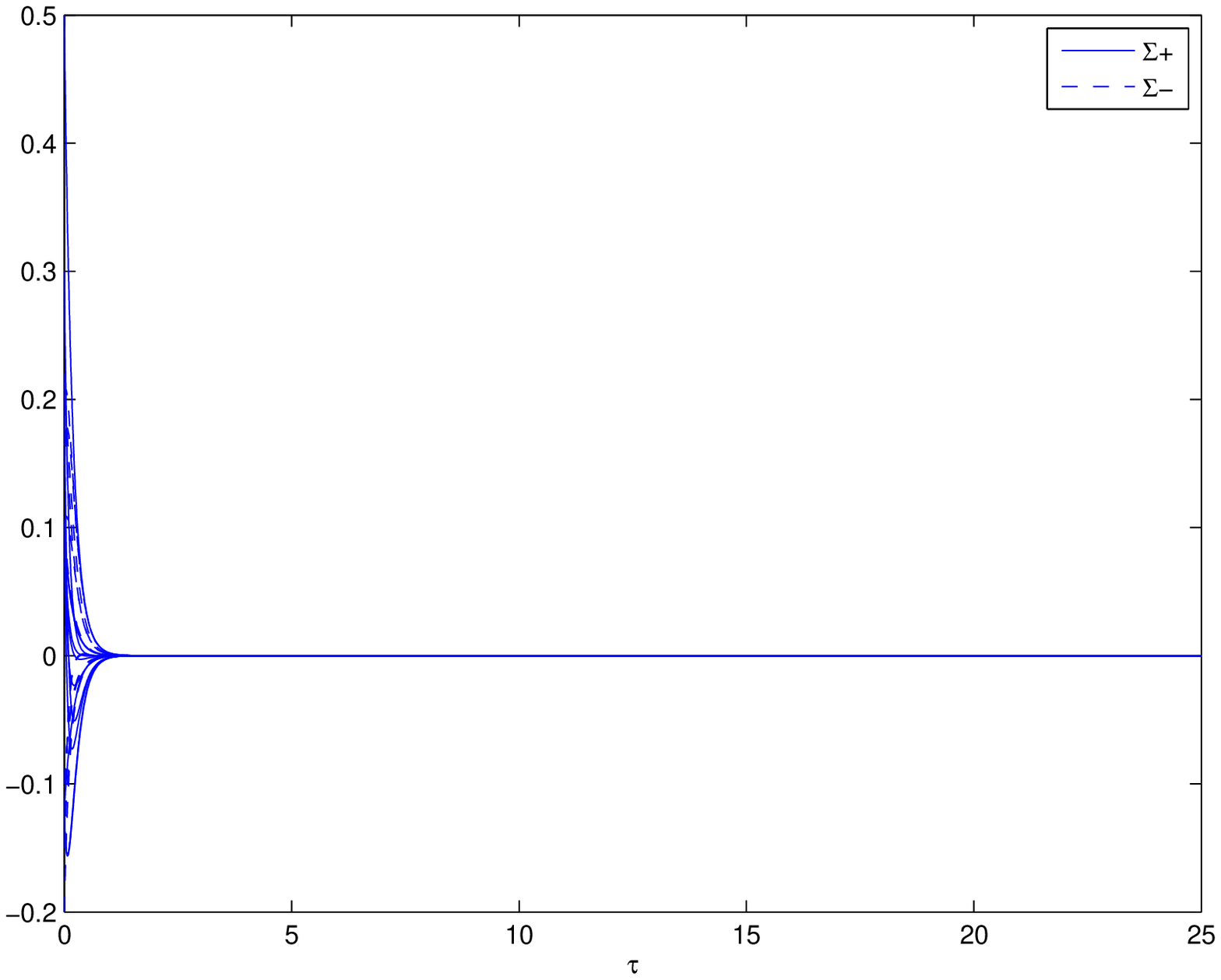}
\end{figure}

\newpage
\subsection{$\xi_{0} = 0.30$, $\eta_{0} = 1 $, $w = 0.325$ (Dust/Radiation Mixture)}
\begin{figure}[h]
\label{fig:fig2}
\caption{This figure shows the dynamical system behaviour for $\xi_{0} = 0.30$, $\eta_{0} = 1 $, and $w = 0.325$ . The plus sign indicates the equilibrium point. Notice how the equilibrium point in this case, the Bianchi Type I / Flat FL point is indeed the local sink. The model also isotropizes as can be seen from the last figure, where $\Sigma_{\pm} \to 0$ as $\tau \to \infty$.}
\includegraphics*[scale = 0.45]{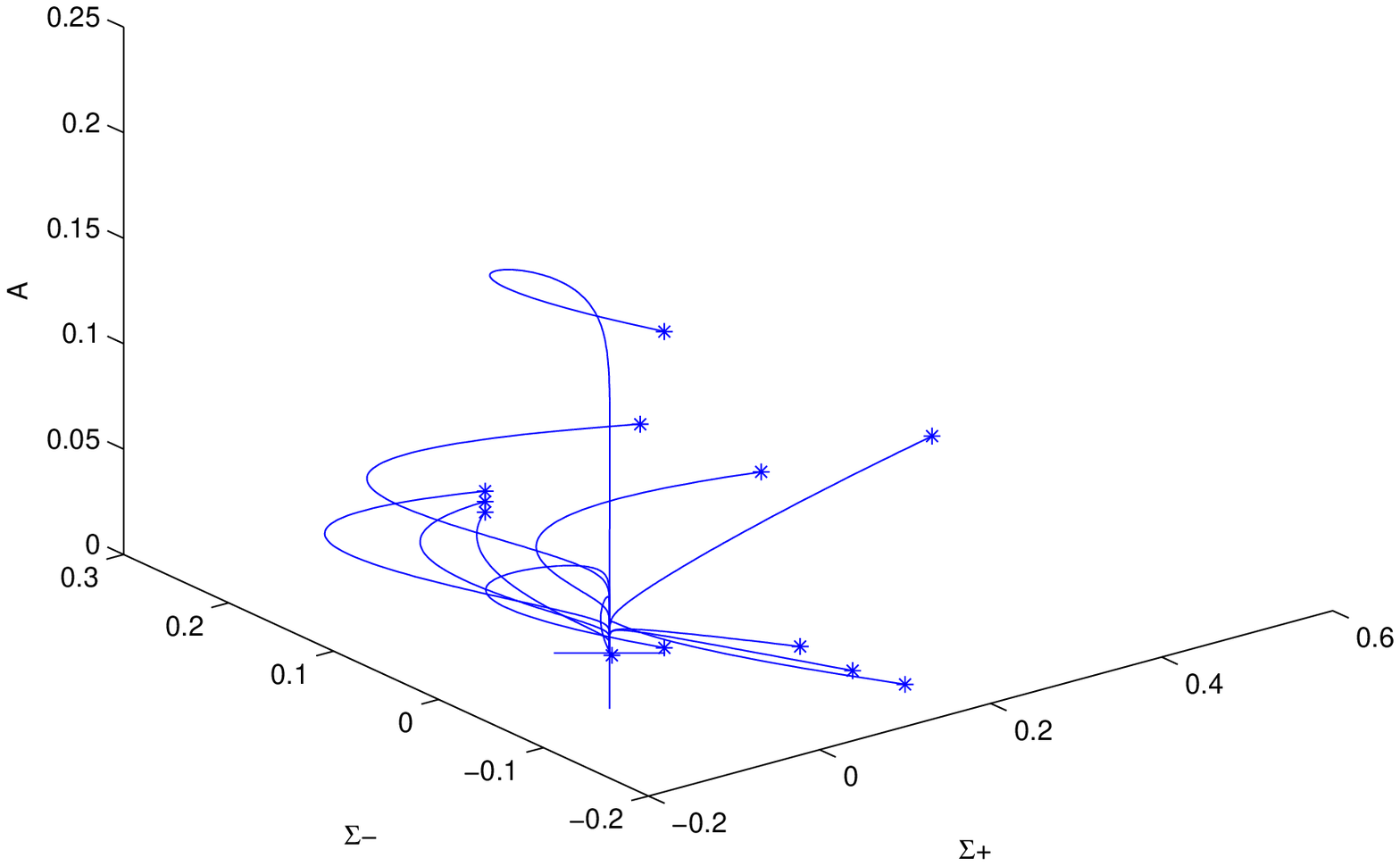}
\includegraphics*[scale = 0.45]{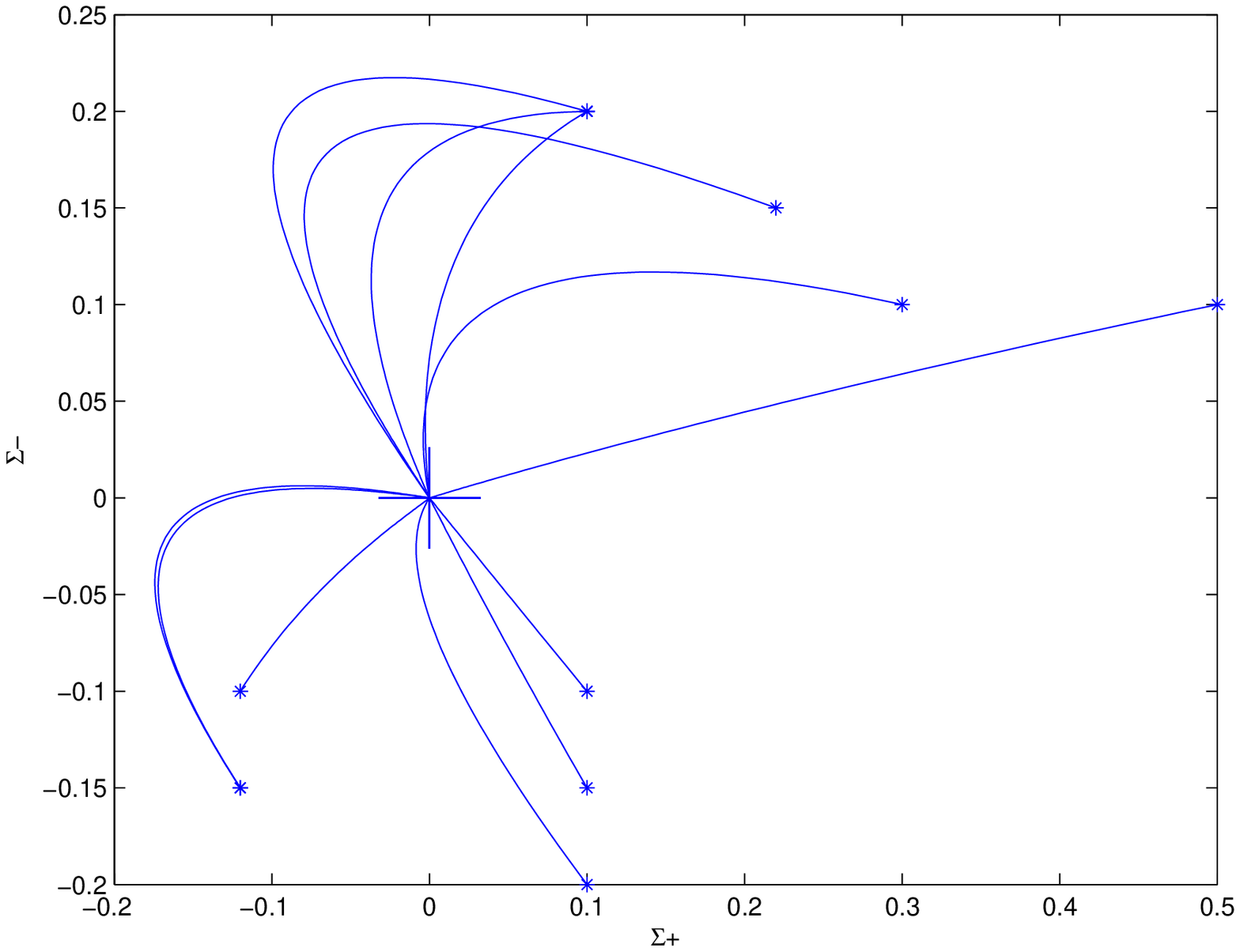}
\linebreak
\linebreak
\includegraphics*[scale = 0.45]{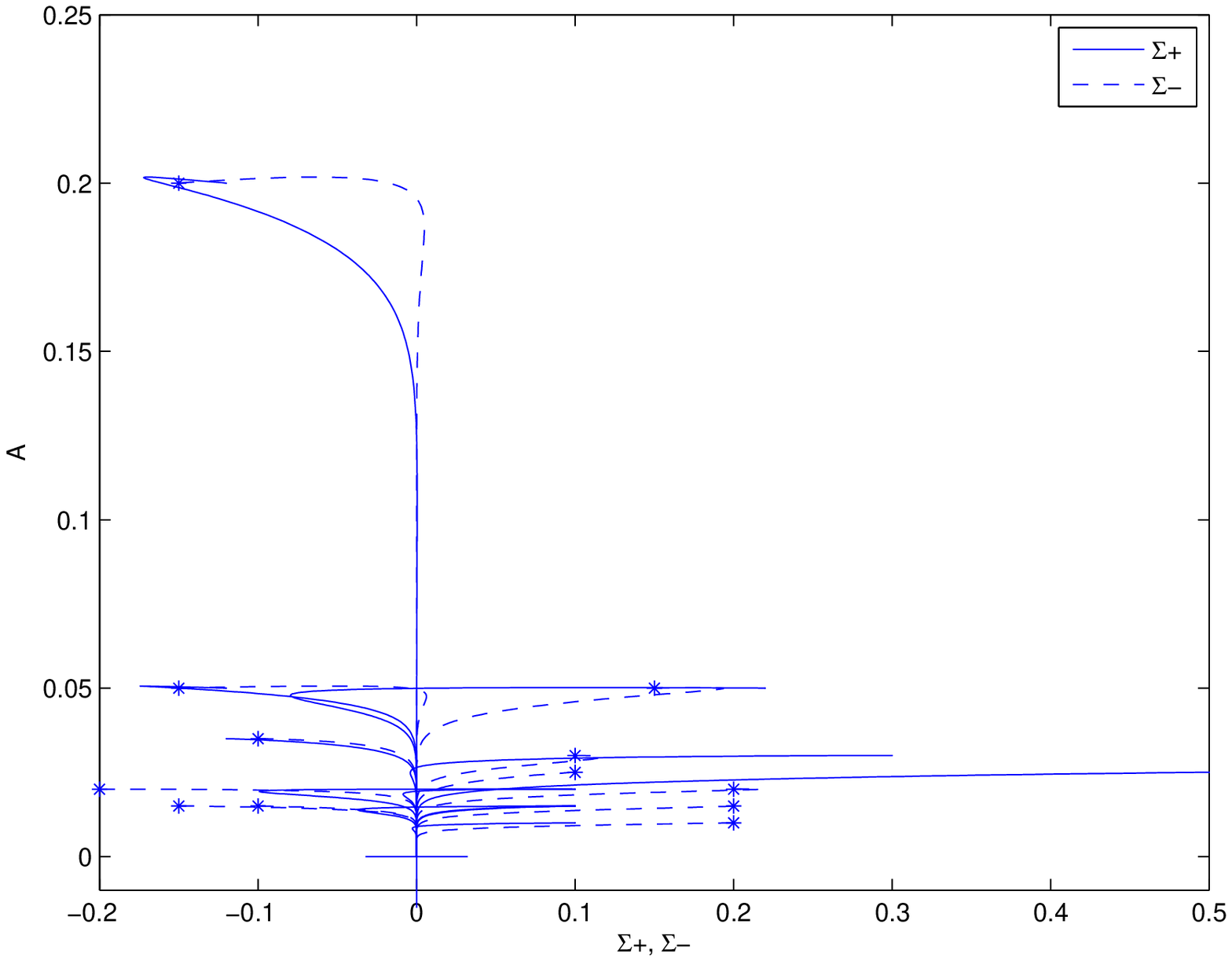}
\includegraphics*[scale = 0.45]{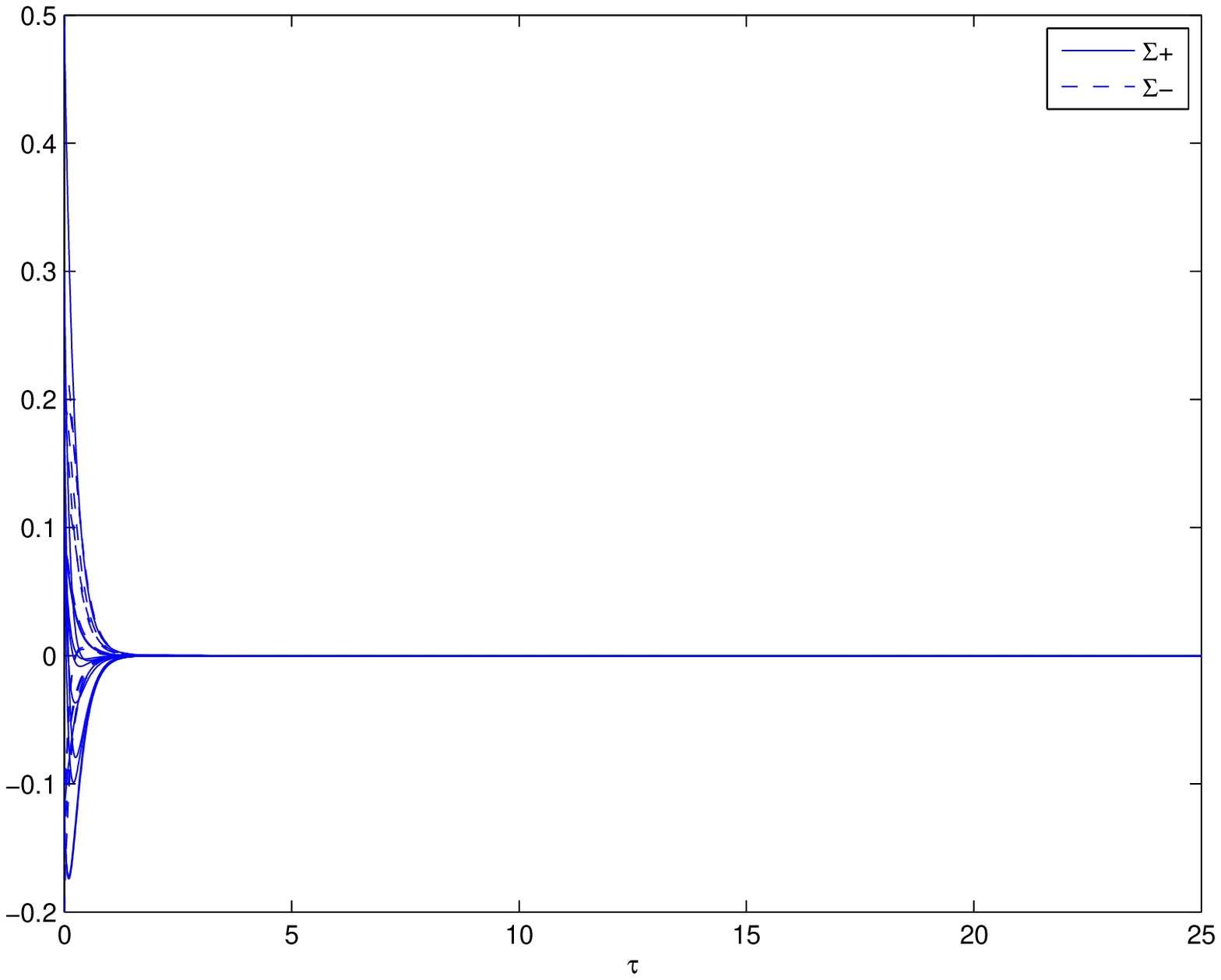}
\end{figure}

\newpage
\subsection{$\xi_{0} = 2 $, $\eta_{0} = \frac{1}{2}$, $w = \frac{1}{3}$ (Radiation)}
\begin{figure}[h]
\label{fig:fig3}
\caption{This figure shows the dynamical system behaviour for $\xi_{0} = 2 $, $\eta_{0} = \frac{1}{2}$, and $w = \frac{1}{3}$  . The plus sign indicates the equilibrium point. Notice how the equilibrium point in this case, the Bianchi Type I / Flat FL point is indeed the local sink. The model also isotropizes as can be seen from the last figure, where $\Sigma_{\pm} \to 0$ as $\tau \to \infty$.}
\includegraphics*[scale = 0.45]{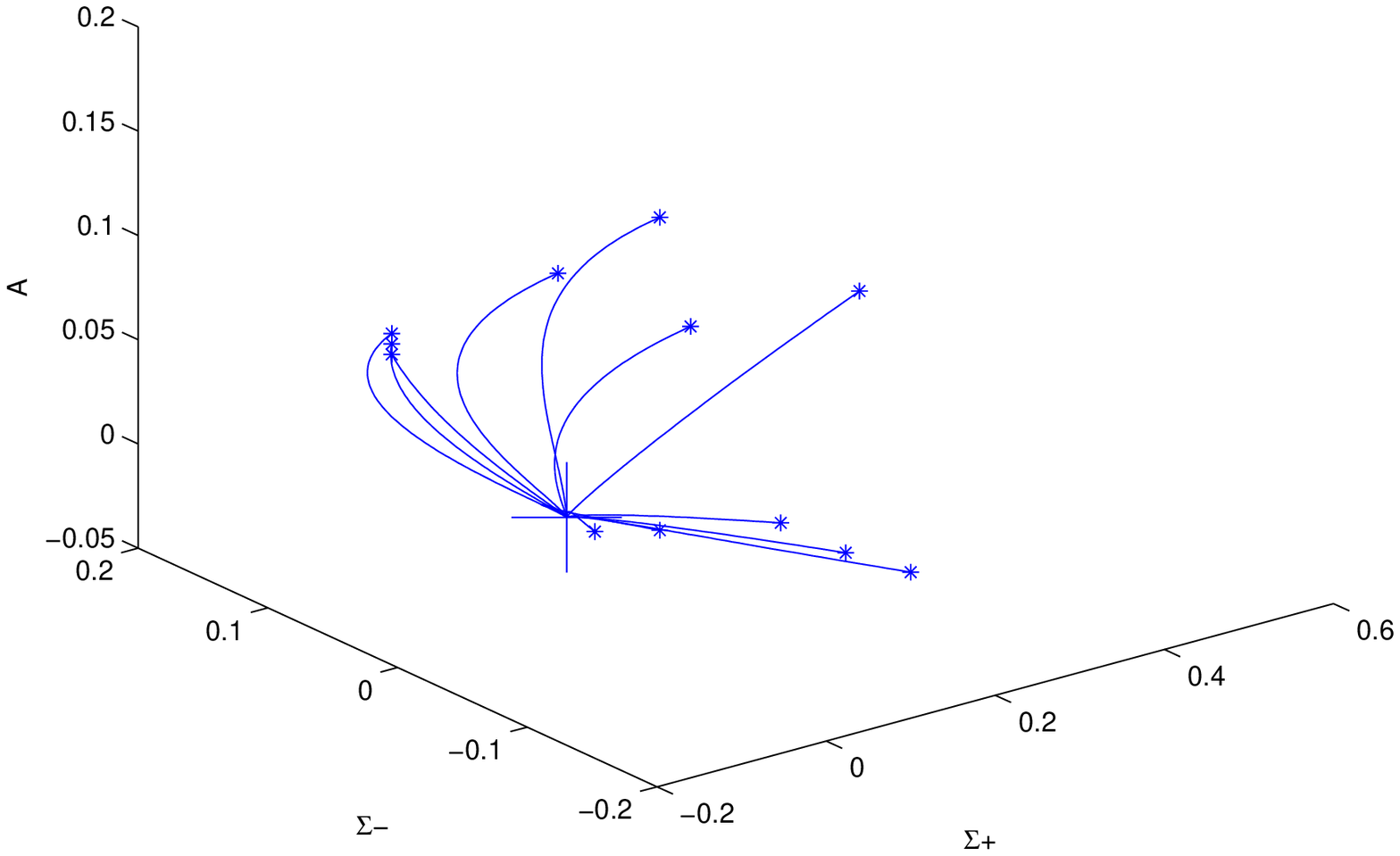}
\includegraphics*[scale = 0.45]{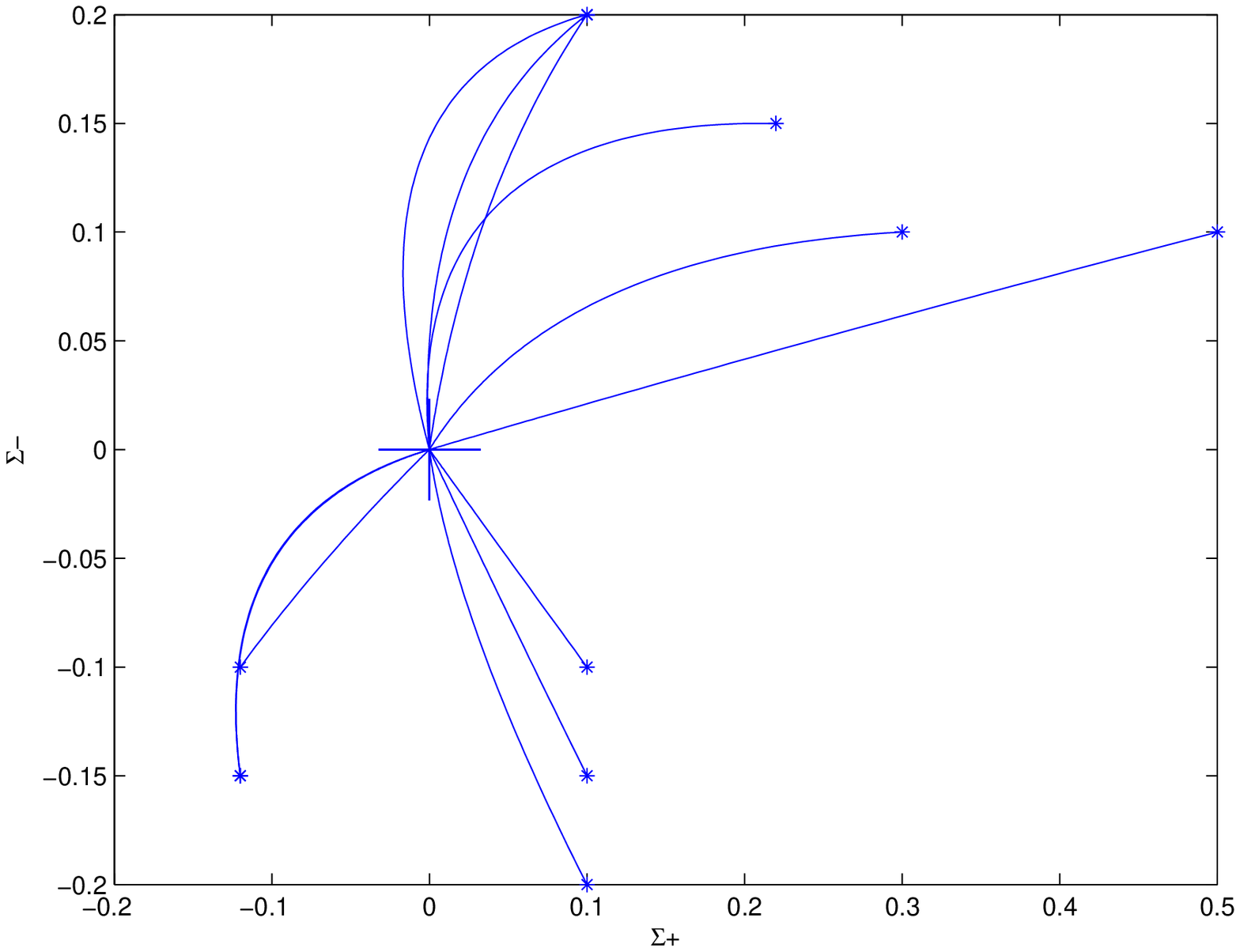}
\linebreak
\linebreak
\includegraphics*[scale = 0.45]{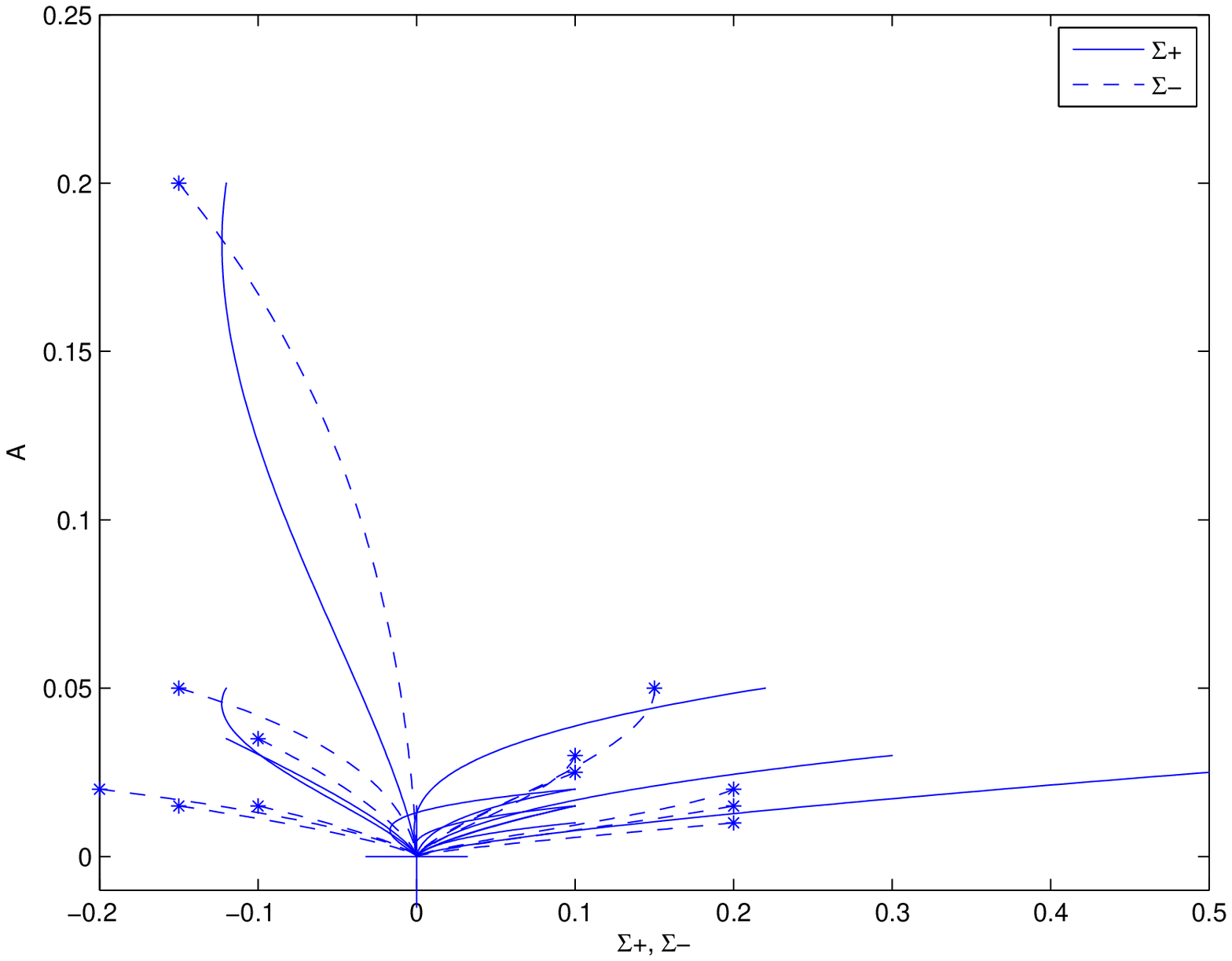}
\includegraphics*[scale = 0.45]{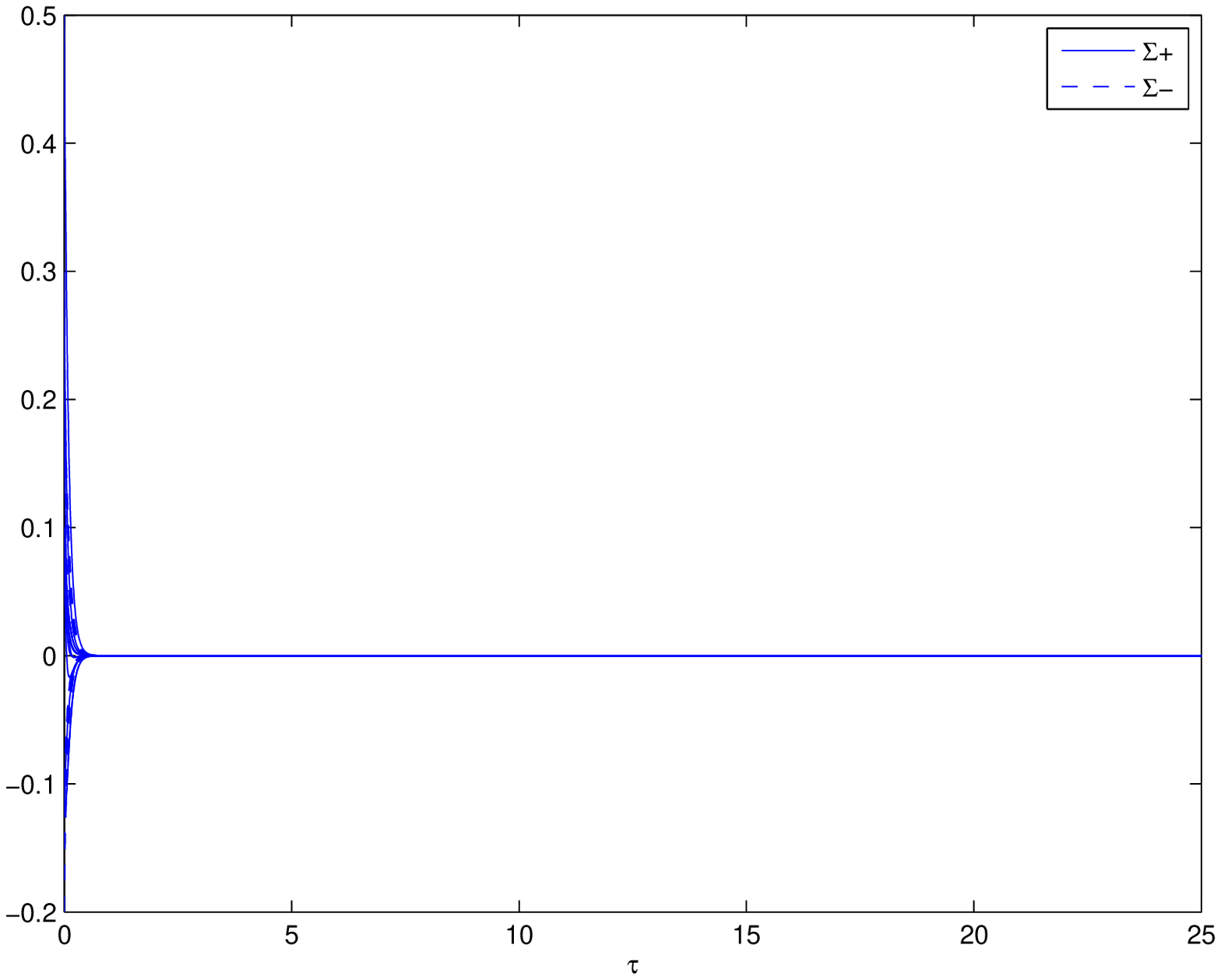}
\end{figure}


\newpage
\subsection{$\xi_{0} = 0.05 $, $\eta_{0} = 1 $, $w = 0$ (Dust)}
\begin{figure}[h]
\label{fig:fig4}
\caption{This figure shows the dynamical system behaviour for $\xi_{0} = 0.05 $, $\eta_{0} = 1$, and $w = 0$. The plus sign indicates the equilibrium point. Notice how the equilibrium point in this case, the Bianchi Type V / Open FL point is indeed the local sink. The model also isotropizes as can be seen from the last figure, where $\Sigma_{\pm} \to 0$ as $\tau \to \infty$.}
\includegraphics*[scale = 0.45]{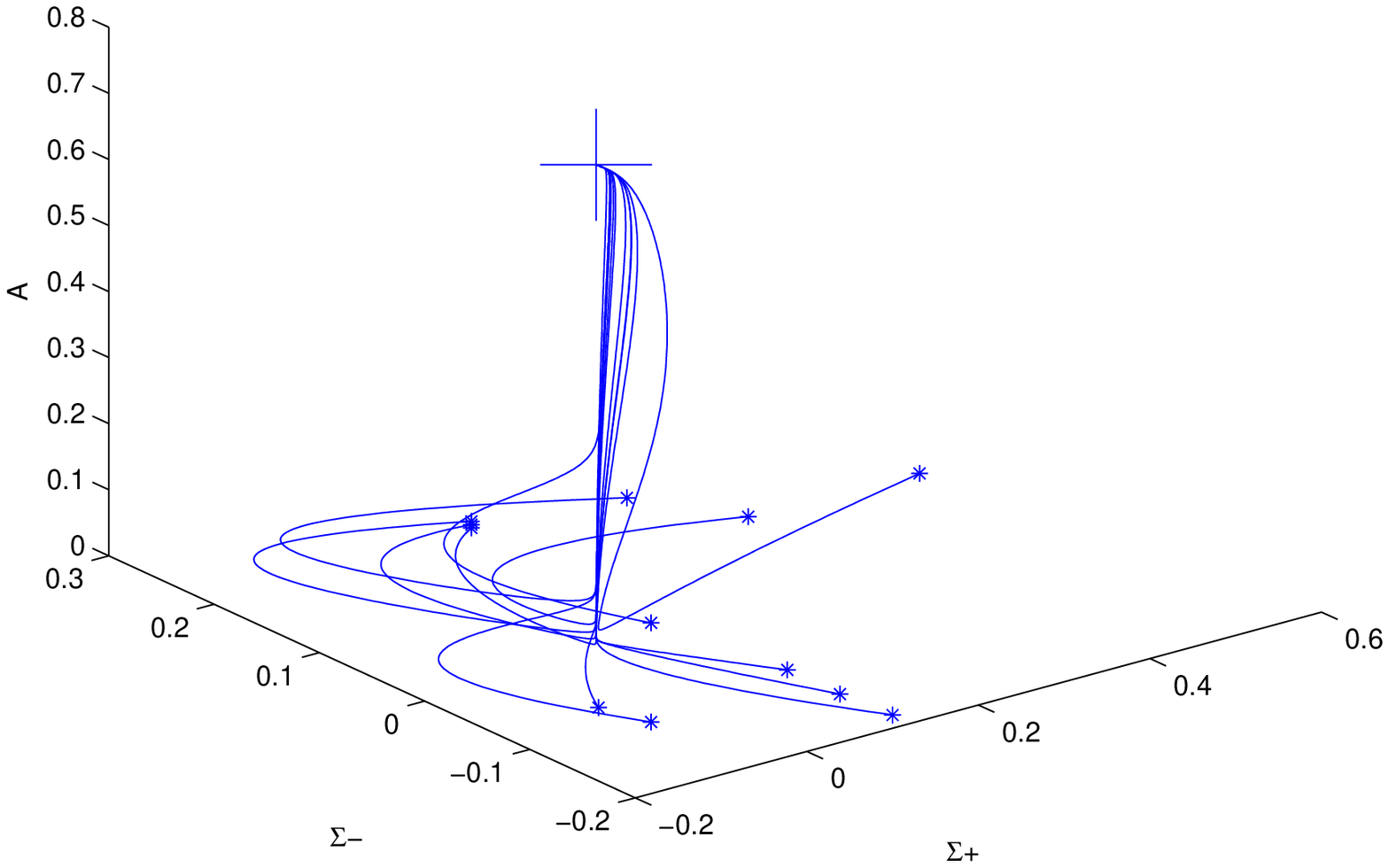}
\includegraphics*[scale = 0.45]{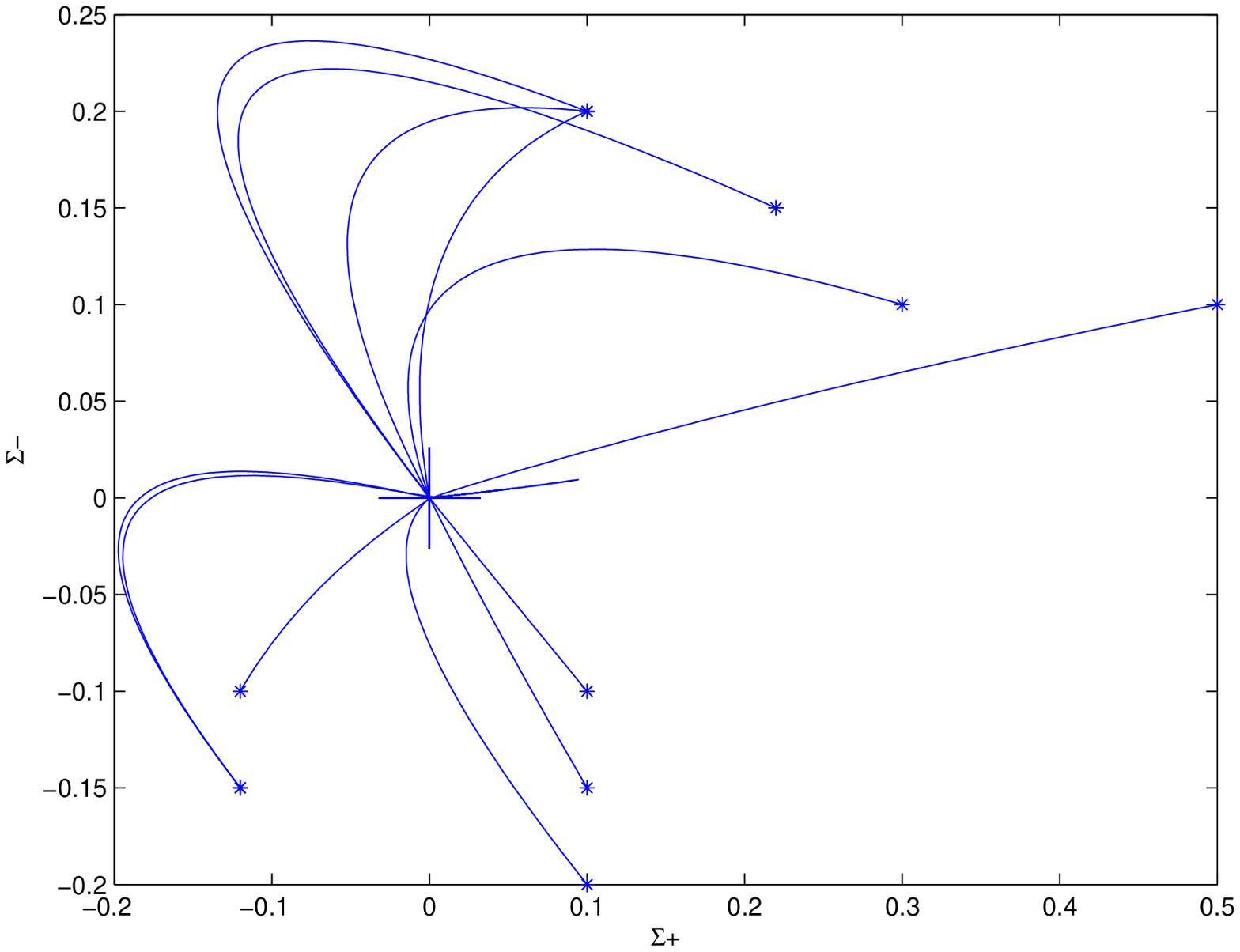}
\linebreak
\linebreak
\includegraphics*[scale = 0.45]{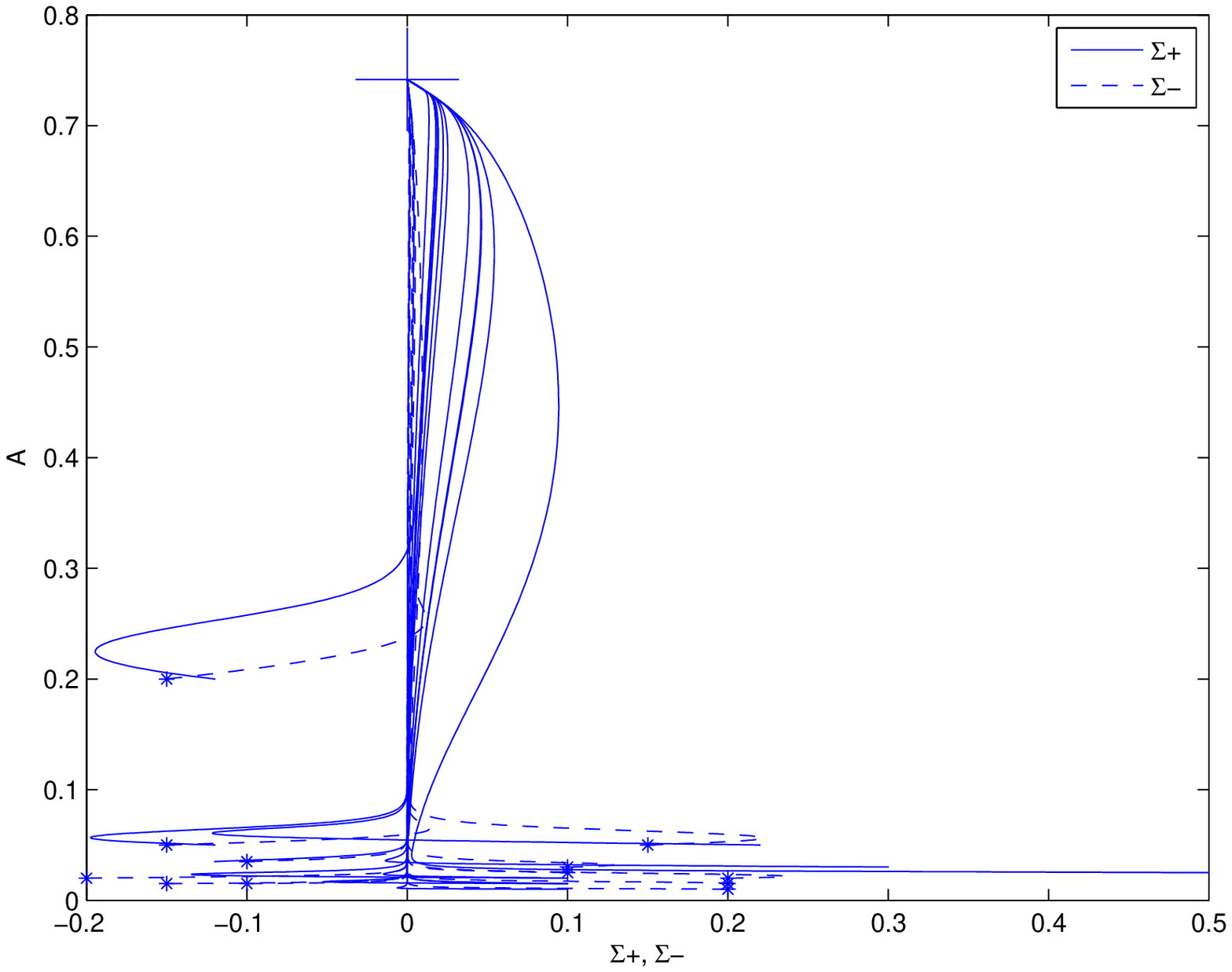}
\includegraphics*[scale = 0.45]{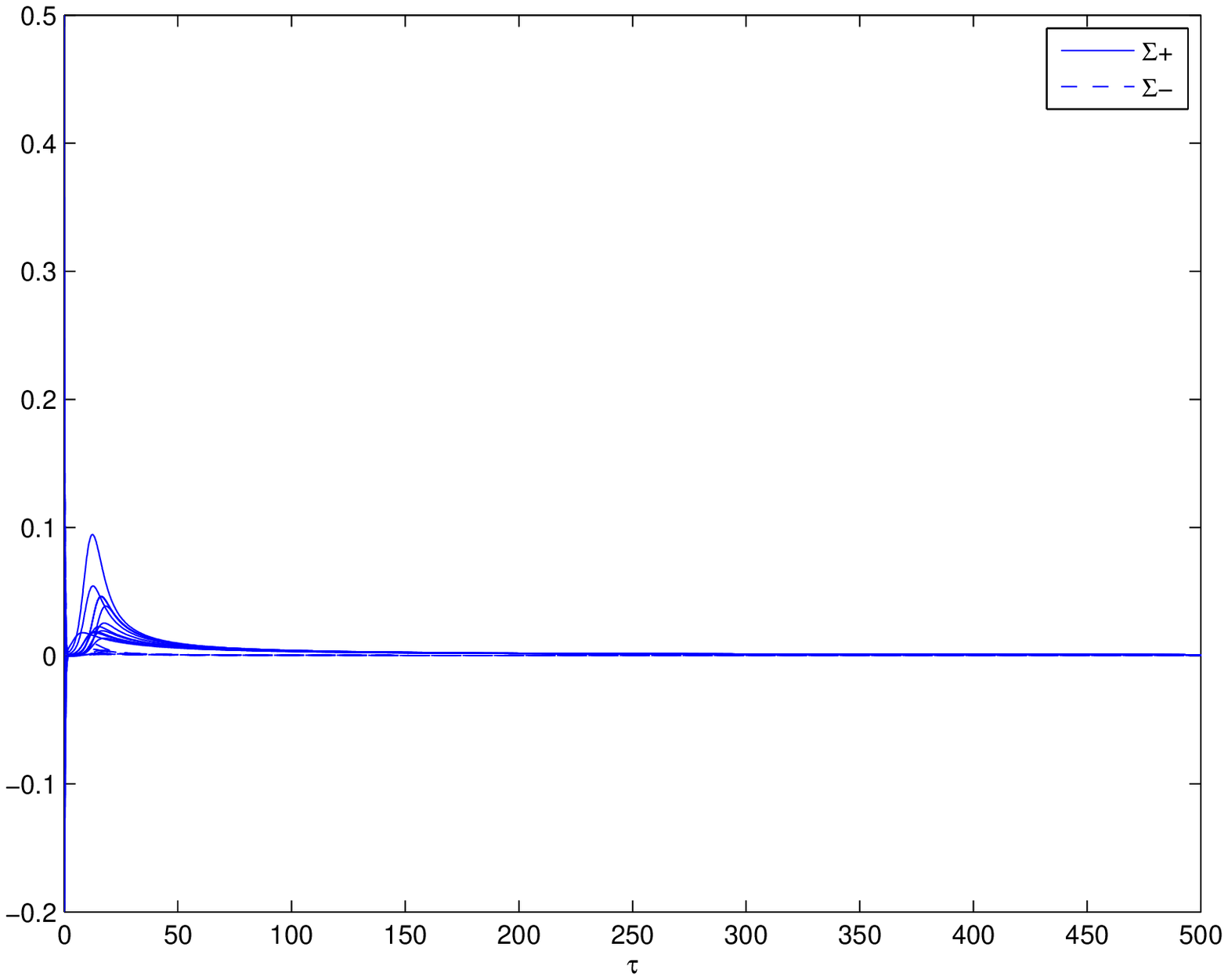}
\end{figure}

\newpage
\subsection{$\xi_{0} = 0.15 $, $\eta_{0} = 1$, $w = 0.325$ (Dust/Radiation Mixture)}
\begin{figure}[h]
\label{fig:fig5}
\caption{This figure shows the dynamical system behaviour for $\xi_{0} = 0.15 $, $\eta_{0} = 1$, and $w = 0.325$. The plus sign indicates the equilibrium point. Notice how the equilibrium point in this case, the Bianchi Type V / Open FL point is indeed the local sink. The model also isotropizes as can be seen from the last figure, where $\Sigma_{\pm} \to 0$ as $\tau \to \infty$.}
\includegraphics*[scale = 0.45]{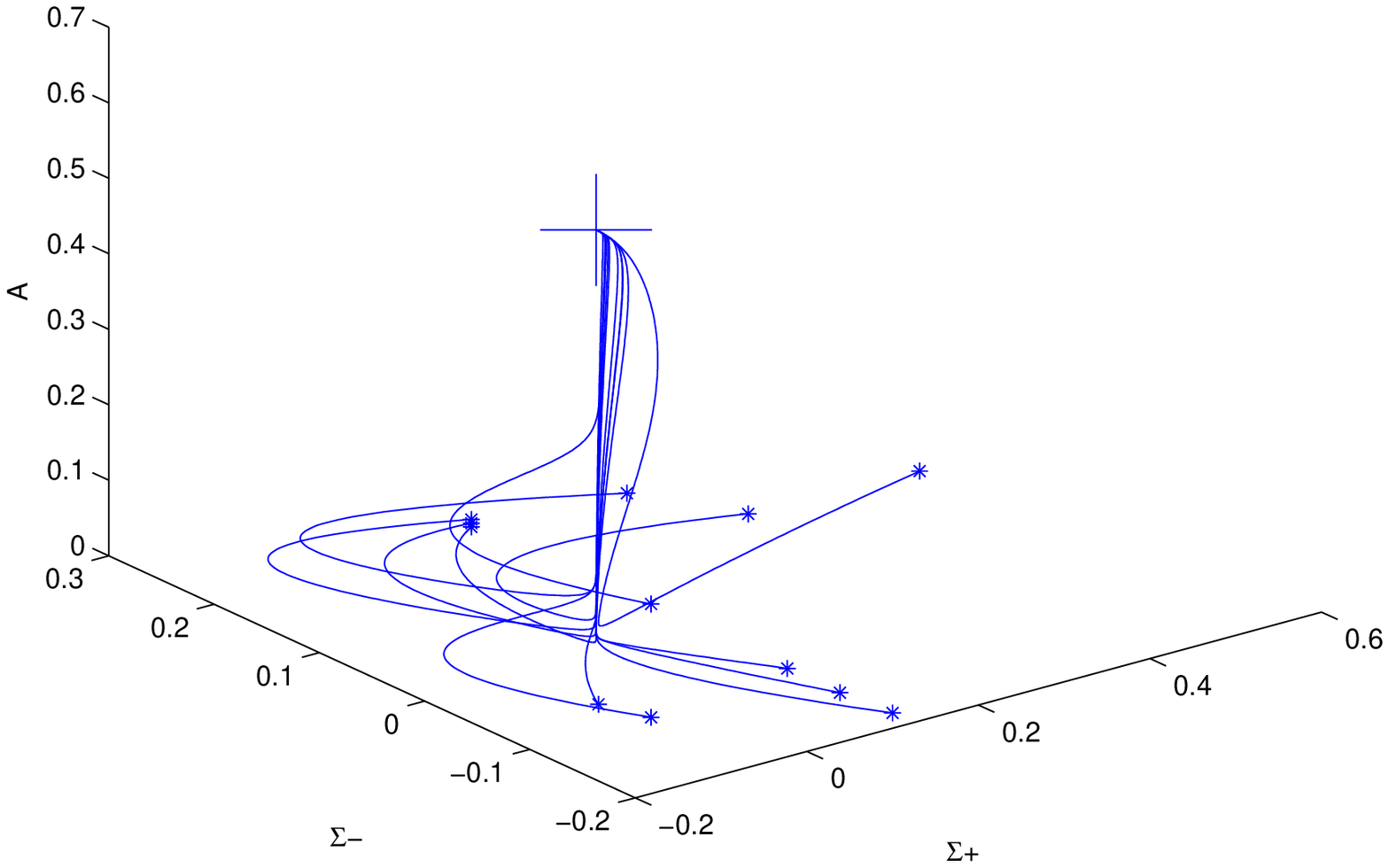}
\includegraphics*[scale = 0.45]{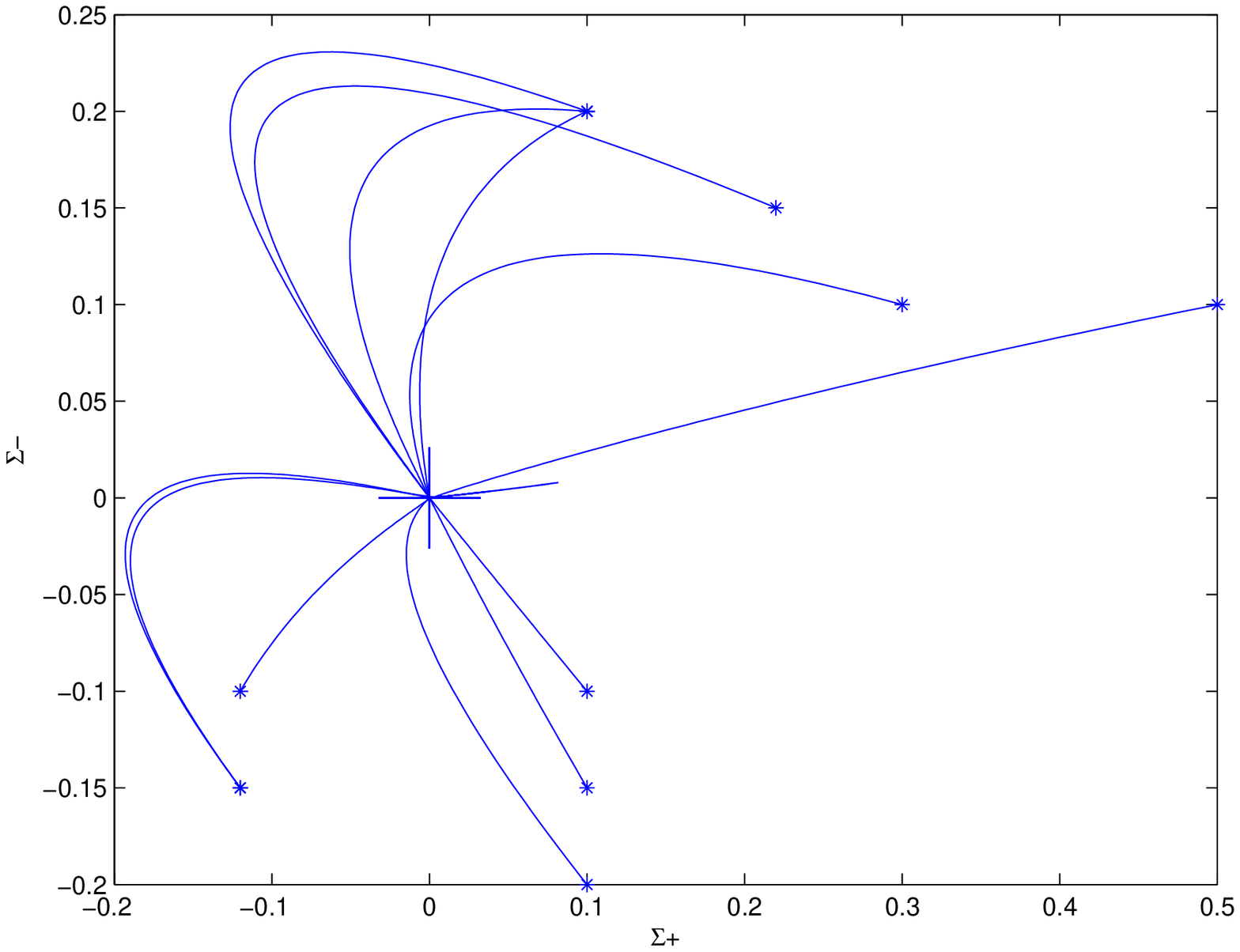}
\linebreak
\linebreak
\includegraphics*[scale = 0.45]{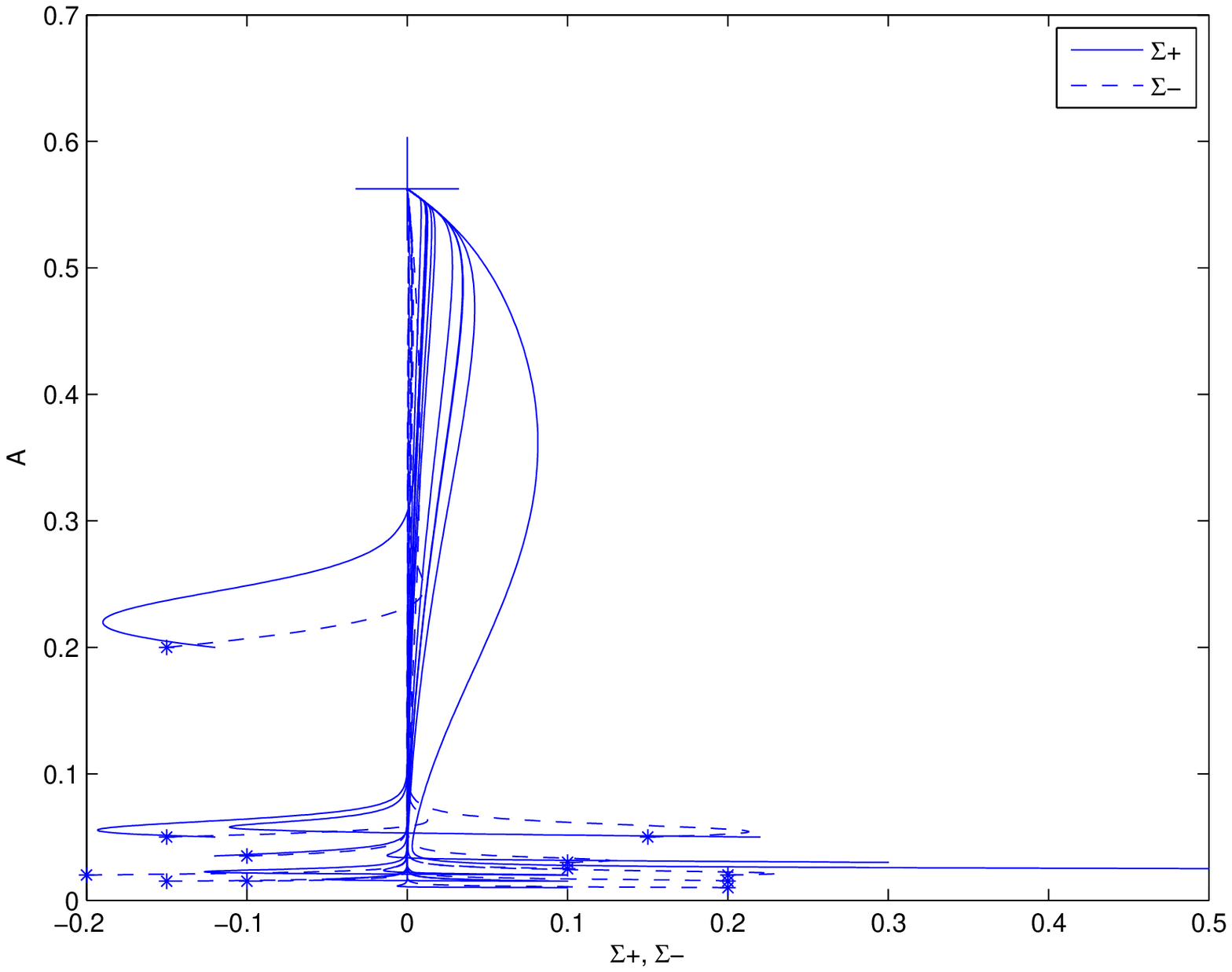}
\includegraphics*[scale = 0.45]{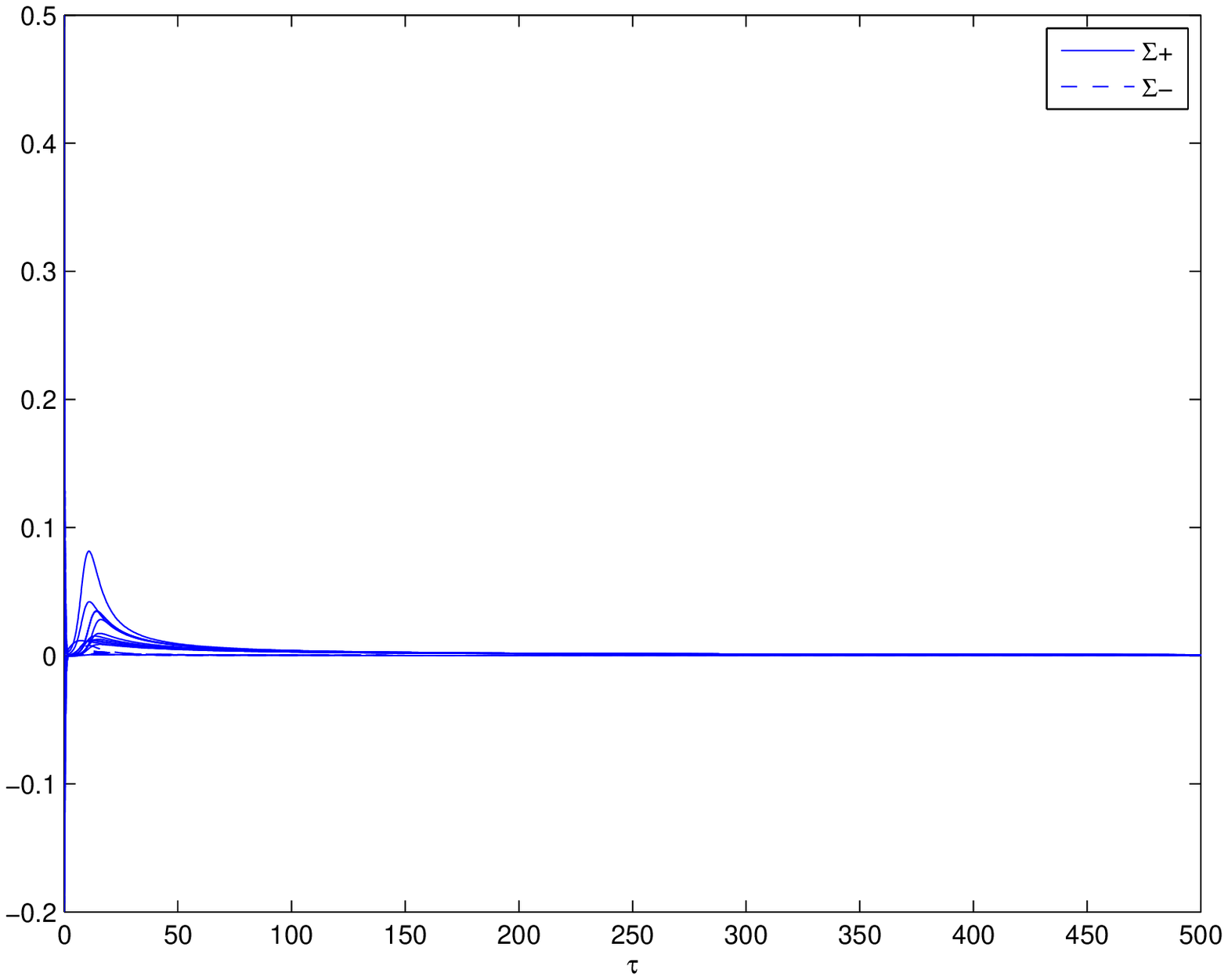}
\end{figure}

\newpage
\subsection{$\xi_{0} = 0 $, $\eta_{0} = 1 $, $w = \frac{1}{3}$ (Radiation)}
\begin{figure}[h]
\label{fig:fig6}
\caption{This figure shows the dynamical system behaviour for $\xi_{0} = 0 $, $\eta_{0} = 1 $, and $w = \frac{1}{3}$ . The plus sign indicates the equilibrium point. Notice how the equilibrium point in this case, the isotropic Milne universe, is indeed the local sink. The model also isotropizes as can be seen from the last figure, where $\Sigma_{\pm} \to 0$ as $\tau \to \infty$.}
\includegraphics*[scale = 0.45]{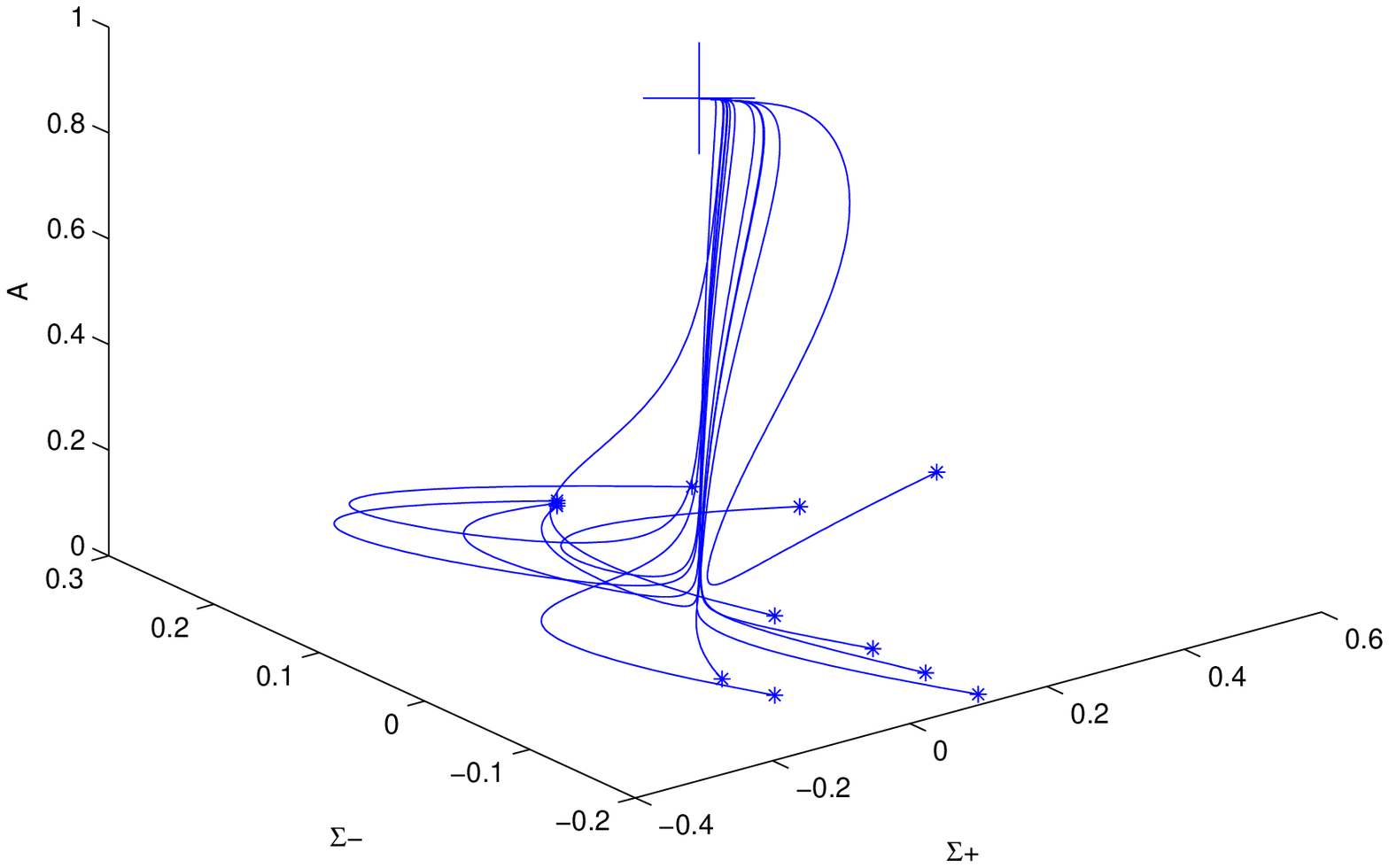}
\includegraphics*[scale = 0.45]{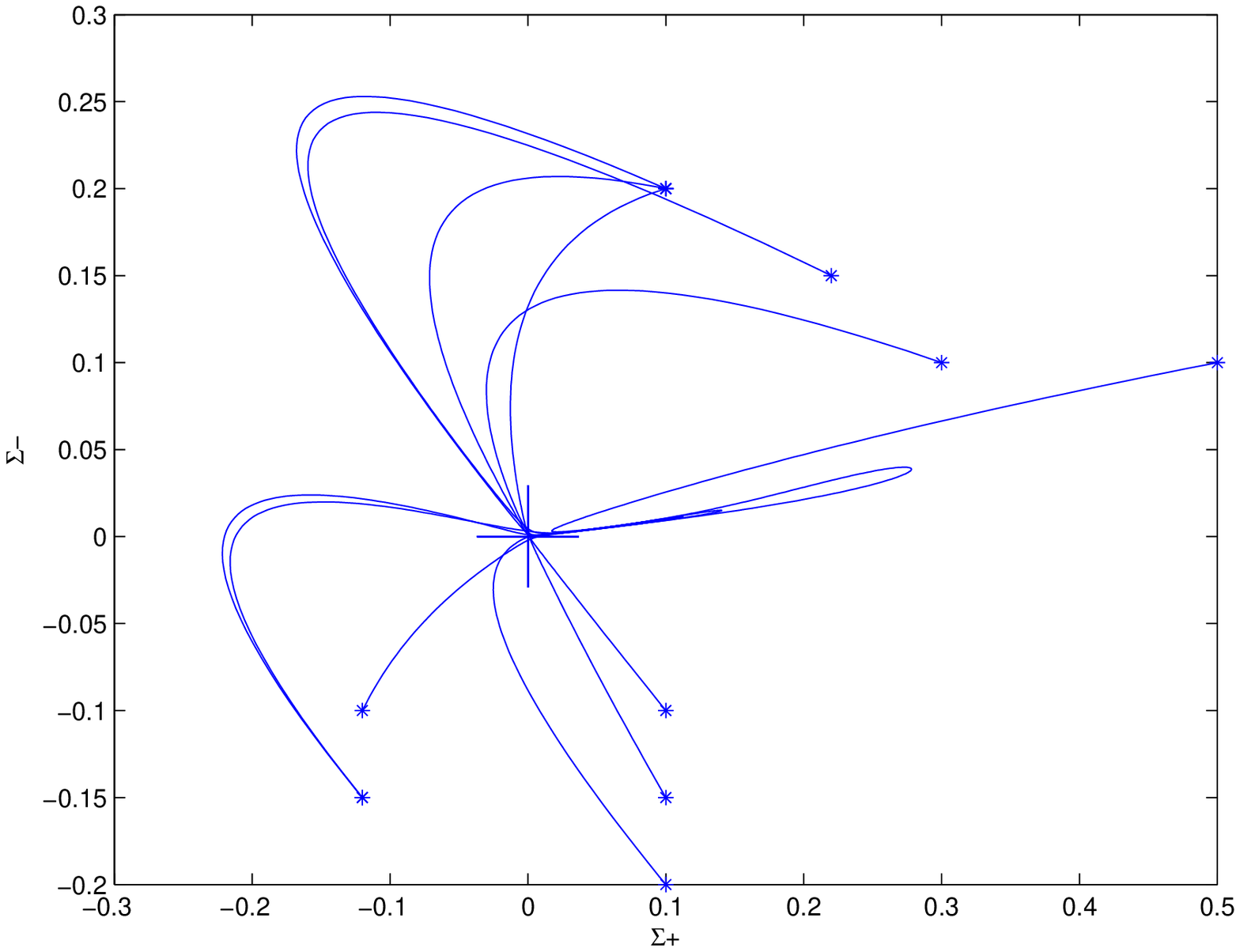}
\linebreak
\linebreak
\includegraphics*[scale = 0.45]{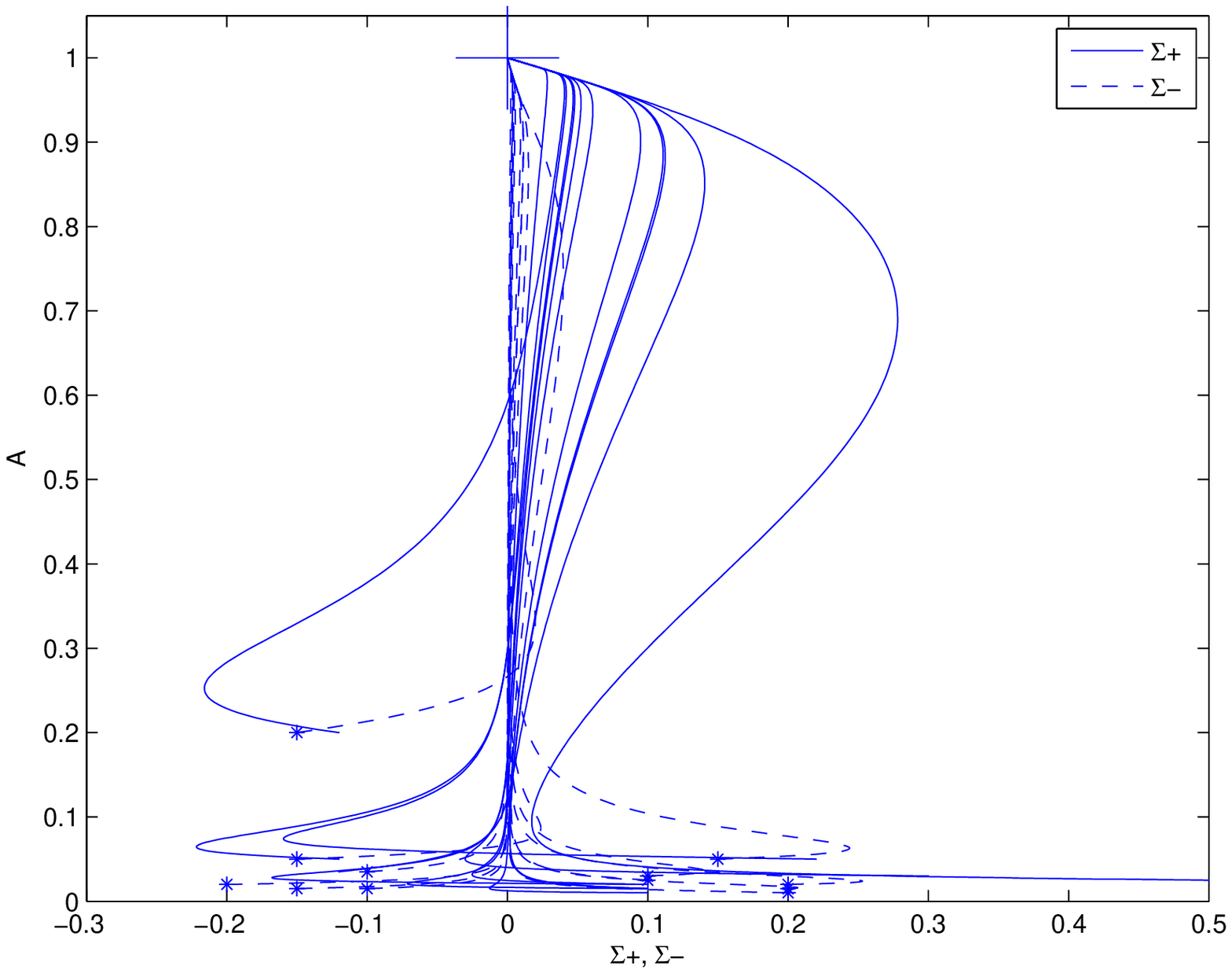}
\includegraphics*[scale = 0.45]{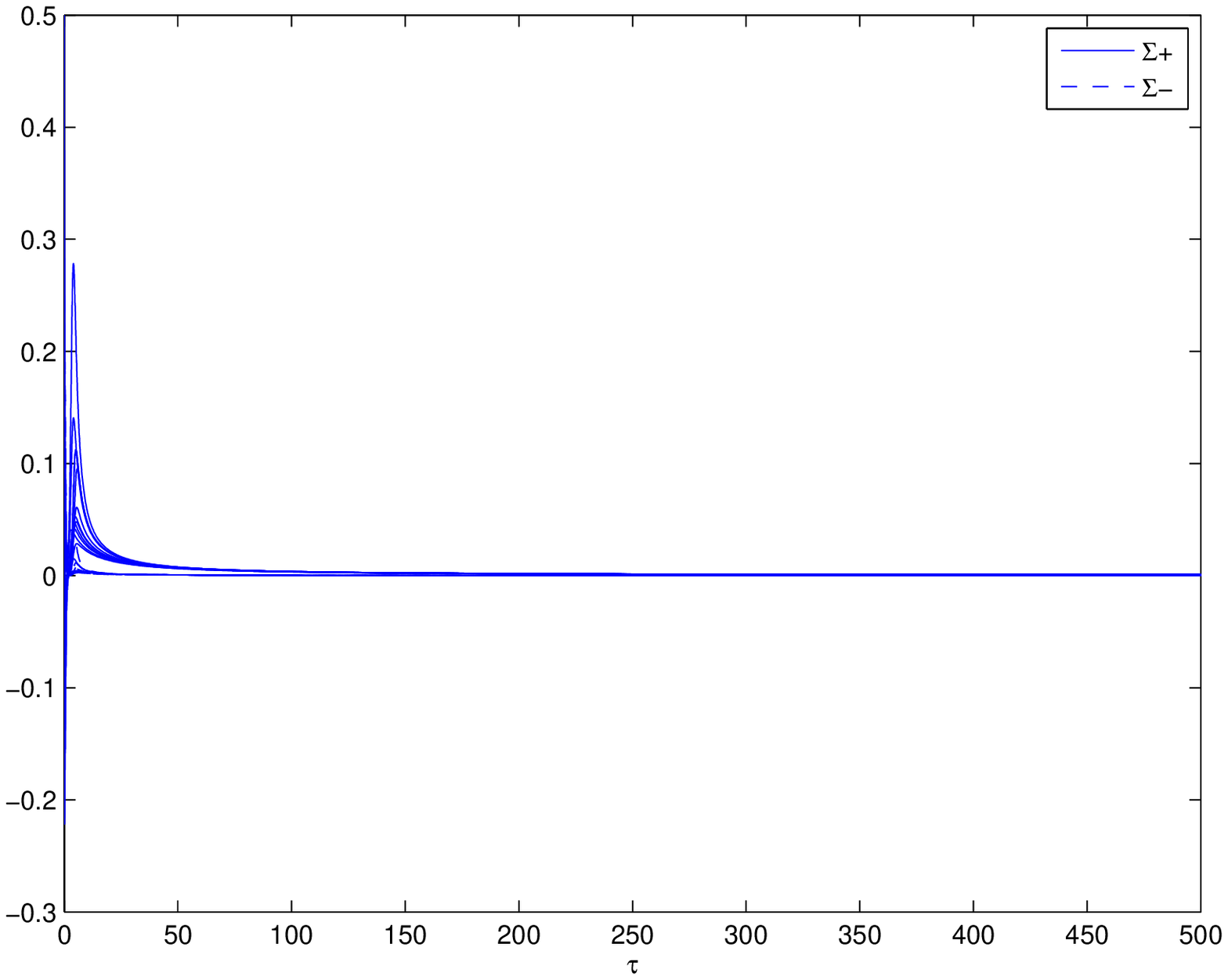}
\end{figure}

%

\subsection{Interpretation of Numerical Results}
In FIGs. (2) ,(3), and (4) we chose values for $\xi_{0}$, $\eta_{0}$, and $w$ that satisfied  Eq. (\ref{eq:restr1}), such that we could model physically interesting situations of dust, a radiation/dust mixture, and radiation. It was clear from these figures that the dynamical system had a local sink at the origin, corresponding to the flat FL solution. It is also of interest to note that the models isotropized asymptotically, as one would expect from any models that asymptotically approach the FL solutions.

In FIGs. (5), (6), and (7) we chose values for $\xi_{0}$, $\eta_{0}$, and $w$ that satisfied  Eq. (\ref{eq:restrV}), such that we could model the physically interesting situations of dust, a radiation/dust mixture, and radiation. It is clear from the numerical simulations which were done over sufficiently long time scales that the models had the Bianchi Type V / open FL solution as a local sink. The models also were found to asymptotically isotropize. Interestingly, in FIG. (7), where we chose $\xi_{0} = 0$, which according to Eq. (\ref{eq:restrV}), then set $A = 1$, the equilibrium point represented the isotropic Milne universe. As one can see from the numerical simulations, it is clear that isotropic Milne universe is a local sink, which we will elaborate on further in what follows.

In their detailed study of the asymptotic behaviour of Bianchi Type Class B models, Hewitt and Wainwright \cite{hewwain} state two interesting conjectures that apply to our present work. Both conjectures have to do with asymptotic stability in a global sense. The first conjecture states that a Bianchi Type IV perfect fluid model with equation of state parameter $w$ satisfying $-1 < w < -\frac{1}{3}$, is asymptotic at late times to the flat FL model. Looking at our inequality in Eq. (\ref{eq:restr1}), we said that the viscous fluid Bianchi Type IV model has the flat FL solution as a local sink if  $0 \leq \xi_{0} \leq \frac{4}{9}$ and $-1 \leq w <\frac{1}{3} \left(-1 + 9 \xi_{0}\right)$. It is our assumption that Hewitt and Wainwright's conjecture only considered \emph{inviscid} perfect fluid models, such that $\xi_{0} = \eta_{0} = 0$, as it is well-known that a perfect fluid can indeed include a bulk viscous pressure. If one sets $\xi_{0} = \eta_{0} = 0$, the inequality in Eq. (\ref{eq:restr1}) becomes $-1 \leq w < -\frac{1}{3}$, which would match the findings of Hewitt and Wainwright. Of course, the second inequality in Eq. (\ref{eq:restr1}) where $\xi_{0} > \frac{4}{9}$ is not considered by Hewitt and Wainwright, but the flat FL model is a local sink in this region as well.

Hewitt and Wainwright's second conjecture is that a Bianchi Type IV perfect fluid model with equation of state parameter $w$ satisfying $-\frac{1}{3} < w < 1$ is asymptotic at late times to a plane-wave model. Once again, setting $\eta_{0} = \xi_{0} = 0$ in Eq. (\ref{eq:restrV}), the inequality reduces to $-\frac{1}{3} < w < 1$, which is precisely the domain under consideration and Hewitt and Wainwright's work. Our model, according to Eq. (\ref{eq:restrV}) has a Bianchi Type V model as a local sink in this domain. Interestingly, the local sink represents an open FL model if $0 < A < 1$ which is true if $0 < \xi_{0} < \frac{4}{9}$, and represents the Milne model if $\xi_{0} = 0$, where $A = 1$.  The Milne model is the isotropic limit of the Bianchi Type IV plane wave solutions \cite{hervikvan}, and so Hewitt and Wainwright's second conjecture is satisfied in this case as well.

Therefore, there is strong evidence that Hewitt and Wainwright's conjectures for perfect, inviscid fluids can be extended to models having viscous fluids, at least for the viscous fluids considered in this paper with constant expansion-normalized bulk and shear viscosity coefficients. Indeed, looking at the auxiliary equation for $\Omega'$ in Eqs. (\ref{eq:evolutionsys1}), we obtain
\begin{equation}
\Omega' = (2q-1)\Omega - 3w\Omega + 9\xi_{0} + \frac{2}{3}\eta_{0} \left[3 \Sigma_{3}^2 + \frac{9}{2}\Sigma_{-}^2 + \frac{1}{2}\Sigma_{+}^2\right].
\end{equation}
Applying Eq. (\ref{eq:q1}), we see that our equation for $\Omega'$ becomes
\begin{equation}
\label{eq:omegaprime}
\Omega' = \frac{1}{3} \left[ \eta_{0}\left(6 \Sigma_{3}^2 + 9 \Sigma_{-}^2 + \Sigma_{+}^2\right) + 27\xi_{0} + \Omega \left(-3 + 12 \Sigma_{3}^2 + 12 \Sigma_{-}^2 + 12\Sigma_{+}^2 - 9w - 27\xi_{0}\right) + \Omega^{2} \left(3 + 9 w\right)\right].
\end{equation}
Let us consider for the time being inflationary models where $-1 \leq w < -\frac{1}{3}$ and $\Omega > 0$ as is done in \cite{hewwain}. We will also make the general assumption that $\xi_{0} \geq 0$, $\eta_{0} \geq 0$. We see from Eq. (\ref{eq:omegaprime}) that
\begin{equation}
\label{eq:omegaprime2}
\frac{1}{3} \left[ \eta_{0}\left(6 \Sigma_{3}^2 + 9 \Sigma_{-}^2 + \Sigma_{+}^2\right) + 27\xi_{0} + \Omega \left(-3 + 12 \Sigma_{3}^2 + 12 \Sigma_{-}^2 + 12\Sigma_{+}^2 - 9w - 27\xi_{0}\right) + \Omega^{2} \left(3 + 9 w\right)\right] = 0,
\end{equation}
if  $\Omega = 1$. We therefore can extend Hewitt and Wainwright's conjecture for inflationary models as follows. If $-1 \leq w < \frac{1}{3}$, $\xi_{0} \geq 0$, $\eta_{0} \geq 0$, and $\Omega > 0$, then for any orbit $\Gamma$, $\omega(\Gamma) = P(I)$, where $P(I)$ characterizes the flat FL point.  We see from the calculation above that the right side of Eq. (\ref{eq:omegaprime}) vanishes if $\Omega = 1$, which is precisely the conclusion reached by Hewitt and Wainwright for inviscid perfect fluids. Therefore, by the LaSalle invariance principle, $\omega(\Gamma) \subseteq \left\{\Omega = 1\right\}$. Since $\Omega$ was assumed to be strictly increasing and $\Omega = 1$ denotes $P(I)$, it follows that the non-vacuum Bianchi IV model under consideration here is asymptotic in the future to the flat FL model and hence, isotropizes.

For the case where $-\frac{1}{3} < w < 1$, the proof of the existence of asymptotic sinks is much more difficult. All we have been able to do is provide some strong evidence for a local sink through our computations of the eigenvalues in Eqs. (\ref{eq:restr4}), (\ref{eq:eigV}), (\ref{eq:restrV}), which is further supported by the long-time numerical solutions presented in FIGs. (5), (6) and (7). 

As mentioned in the introduction, Hervik, van den Hoogen, and Coley studied the future asymptotic behaviour of titled Bianchi Type IV models with a perfect fluid. We therefore find it appropriate to compare our results to theirs in the limits of no viscosity and tilt. In this regard, they also found as equilibrium points: the Bianchi Type I/flat FL solution, the Bianchi Type V / open FL solution, and as a special case of the Bianchi Type IV, the isotropic Milne solution. Indeed, they also found that for $-\frac{1}{3} < w < 1$, the isotropic Milne solution is a stable future attractor in the isotropic limit of the plane-wave equilibrium points of Bianchi Type IV. They also confirmed that for inflationary fluids, where $-1 < w < -\frac{1}{3}$, the flat Friedmann solution is indeed a stable future attractor \cite{hervikvan}, which was also a conclusion reached by Hewitt, Bridson, and Wainwright in their study of the titled Bianchi Type II models \cite{hewittbridsonwainwright}. 
\section{Conclusions}
We have used a dynamical systems approach combined with a sophisticated numerical analysis to analyze the future asymptotic behaviour of a non-tilted Bianchi Type IV viscous fluid model with constant nonnegative expansion-normalized shear and bulk viscosity coefficients. After deriving the equations of motion, we proceeded with a  fixed-point analysis and found the corresponding equilibrium points. The future asymptotic behaviour of the non-titled Bianchi IV viscous fluid models can be summarized as follows:
\begin{enumerate}
\item $\eta_{0} \geq 0, \quad \left\{\left[0 \leq \xi_{0} \leq \frac{4}{9}\right] \wedge \left[-1 \leq w < \frac{1}{3}\left(-1 + 9 \xi_{0}\right)\right]\right\} \vee \left\{\left[\xi_{0} > \frac{4}{9}\right] \wedge \left[ -1 \leq w < 1\right] \right\}$: Asymptotically flat FL.
\item $\eta_{0} \geq 0, \quad \left\{0 < \xi_{0} < \frac{4}{9}\right\} \wedge \left\{\frac{1}{3} \left[-1 + 9 \xi_{0}\right] < w < 1\right\}$: Asymptotically open FL.
\item $\eta_{0} \geq 0 , \quad \left\{\xi_{0} = 0 \right\} \wedge \left\{-\frac{1}{3} < w < 1\right\}$: Asymptotically isotropic Milne universe (which is an isotropic limit of the plane-wave equilibrium points of Bianchi Type IV \cite{hervikvan}).
\end{enumerate}

We also established that bifurcations exist such that the spatial curvature destabilizes the flat FL point if $\xi_{0} = \frac{1}{9} \left(1 + 3w\right)$, and destabilizes the Bianchi Type V point if $\xi_{0} = \frac{1}{9} \left(1 + 3w\right)$ and $-\frac{1}{3} < w < 1$. 

Finally, we showed that for each numerical solution, our Bianchi Type IV model with constant viscous coefficients isotropized at late times for the regions corresponding to the flat FL, open FL, and isotropic Milne equilibrium points. 

\section{Acknowledgements}
The authors would like to thank the reviewer for his/her generous and helpful comments. In addition, we would like to thank Charles C. Dyer, George F.R. Ellis, and Sibj\o rn Hervik for helpful comments and clarifications.  The authors gratefully acknowledge the support of the Natural Sciences and Research Council of Canada.

\appendix*
\section{Table of Initial Conditions}
For completeness, we present in this section a table of the initial conditions used in the preceding numerical experiments.
\begin{table}[h]
\begin{center}
\begin{tabular}{|c|c|}\hline 
\textbf{Initial Conditions} & $g(\mathbf{x})$ \\\hline 
$\left[\Sigma_{+}, \Sigma_{-}, \Sigma_{3}, N_{1}, A\right] = [0.1, 0.2, -0.386603, 0.01, 0.01]$ & 0  \\\hline 
$\left[\Sigma_{+}, \Sigma_{-}, \Sigma_{3}, N_{1}, A\right] = [0.1, 0.2, -0.773205, 0.01, 0.02]$ & 0 \\\hline 
$\left[\Sigma_{+}, \Sigma_{-}, \Sigma_{3}, N_{1}, A\right] = [0.1, -0.2, 0.426795, 0.01, 0.02]$ & 0 \\\hline 
$\left[\Sigma_{+}, \Sigma_{-}, \Sigma_{3}, N_{1}, A\right] = [0.1, 0.2, -0.579905, 0.01, 0.015]$ & 0 \\\hline 
$\left[\Sigma_{+}, \Sigma_{-}, \Sigma_{3}, N_{1}, A\right] = [0.1,-0.1,0.0475481,0.02, 0.015]$ & 0 \\\hline 
$\left[\Sigma_{+}, \Sigma_{-}, \Sigma_{3}, N_{1}, A\right] = [0.1,-0.15, 0.103798 ,0.02, 0.015]$ & 0 \\\hline 
$\left[\Sigma_{+}, \Sigma_{-}, \Sigma_{3}, N_{1}, A\right] = [0.3,0.1,-0.614711,0.02,0.03]$ & 0 \\\hline 
$\left[\Sigma_{+}, \Sigma_{-}, \Sigma_{3}, N_{1}, A\right] = [0.5, 0.1,-0.145753,0.1,0.025]$& 0 \\\hline 
$\left[\Sigma_{+}, \Sigma_{-}, \Sigma_{3}, N_{1}, A\right] = [0.22,0.15,-0.831051,0.025,0.05]$& 0 \\\hline 
$\left[\Sigma_{+}, \Sigma_{-}, \Sigma_{3}, N_{1}, A\right] = [-0.12,-0.15,0.657846,0.025,0.05]$ & 0 \\\hline 
$\left[\Sigma_{+}, \Sigma_{-}, \Sigma_{3}, N_{1}, A\right] = [-0.12,-0.1,0.177746,0.05,0.035]$ & 0 \\\hline 
\end{tabular} 
\caption{ Initial conditions used in the numerical simulations.}
\end{center}
\label{defaulttable}
\end{table}

\newpage 
\bibliography{sources}

\end{document}